# Analytical Study of Reinforced Concrete Beams Strengthened by FRP Bars Subjected to Impact Loading Conditions

[1]Sajjad Roudsari, [2]Sameer Hamoush, [3]Sayed Soleimani,
[4]Taher Abu-Lebdeh and [5]Mona HaghighiFar

[1]*Department of Computational Science and Engineering,*
*North Carolina A and T State University, 1601 E. Market St., Greensboro, NC, USA*
[2]*Departments of Civil and Architectural Engineering, North Carolina A and T State University, Greensboro, NC, USA*
[3]*Department of Civil Engineering, School of Engineering, Australian College of Kuwait, Kuwait*
[4]*Department of Civil, Architectural and Environmental Engineering,*
*North Carolina A and T State University, Greensboro, NC 27411, USA*
[5]*Department of Structural Engineering, University of Guilan, Guilan, Rasht, Iran*

Corresponding Author:
Taher Abu-Lebdeh
Department of Civil, Architectural and Environmental Engineering, North Carolina A and T State University, Greensboro, NC 27411, USA
Email: taher@ncat.edu

**Abstract:** Civil engineers have considered Fiber Reinforced Polymer (FRP) materials to enhance the performance of structural members subjected to static and dynamic loading conditions. However, there are some design limitations due to uncertainty in the behavior of such strengthened members. This fact is particularly important when considering the complex nature of the nonlinear behavior of materials, the impact loading conditions and geometry of the members having FRP systems. In this research, a new analytical model is developed to analyze structural members strengthened with FRP systems and subjected to impact loading conditions. ABAQUS based finite element code was used to develop the proposed model. The model was validated against nine beams built and tested with various configurations and loading conditions. Three sets of beams were prepared and tested under quasistatic and impact loadings by applying various impact height and Dynamic Explicit loading conditions. The first set consisted of two beams, where one of the beams was reinforced with steel bars and the other was externally reinforced with GFRP sheet. The second set consisted of six beams, with five of the beams were reinforced with steel bars and one of them wrapped by GFRP sheet. The last set was tested to validate the response of concrete beams reinforced by FRP bar. In addition, beams were reinforced with glass and carbon fiber composite bars tested under Quasi-Static and Impact loading conditions. The impact load was simulated by the concept of a drop of a solid hammer from various heights. The numerical results showed that the developed model can be an effective tool to predict the performance of retrofitted beams under dynamic loading condition. Furthermore, the model showed that FRP retrofitting of RC beams subjected to repetitive impact loads can effectively improve their dynamic performance and can slow the progress of damage.

## Introduction

The use of composite sheets and bars can be an effective and usable method for enhancing the structural performance of existing structures when they are subjected to impact loading conditions. Many researches have studied and evaluated the effect of dynamic loads on retrofitted RC structures. Erki and Meier (1999) performed experimental tests on four eight-meter RC beams externally strengthened to enhance the flexural strength. Two beams were retrofitted by CFRP systems and the remaining beams were reinforced by external steel plates. All four beams were tested under impact loadings. The impact load was generated by lifting and dropping a weight from given height into simply supported beams. Results showed that the energy absorption of beam with CFRP laminates is less than that of beams strengthened with external steel plates. White *et al.* (2001) conducted experimental work to investigate the response of RC beams strengthened by CFRP laminates when subjected to high loading rate. They examined nine three-meter long reinforced concrete beams. One beam was a control beam without external reinforcement and the remaining eight beams were externally reinforced with CFRP sheets. Results revealed that beams subjected to rapid loads at a higher rate gained about 5% in strength and in stiffness and energy absorption. They indicated that the change in loading rate did not affect the flexibility and the mode of failure. Tang and Saadatmanesh (2005) performed investigation to evaluate the behavior of concrete beams strengthened with reinforced polymer laminates subjected to impact loadings. Two of the beams were control beams without external reinforcement and the remaining beams were externally reinforced. The results showed that the composite sheets can significantly improve the bending strength and the stiffness of retrofitted RC beams. GoldSton *et al.* (2016) performed experimental investigation on concrete beams reinforced with GFRP bars under static and impact loading. In their work, they performed experimental tests on twelve reinforced concrete beams. The focus was to evaluate the effect of glass fiber reinforcement on the strength of the concrete beam when they are under static and dynamic impact loading conditions. Six of the tested beams were reinforced with GFRP bars and subjected to static loading and the remaining six were reinforced externally with GFRP systems. They showed that the higher GFRP reinforcement ratio resulted in higher rate of cracking and less ductility under static loading conditions. But under dynamic loads, the beams' strength was 15-20% higher than the strength obtained by the static loading conditions. Liao *et al.* (2017) conducted experimental studies and numerical simulation to evaluate the behavior of RC beams retrofitted with High Strength Steel Wire Mesh and High-Performance Mortar (HSSWMHPM) under impact loads. The results of both laboratory samples and finite element analysis showed a significantly improvement in the impact resistance as well as an improvement in the ductility of beams reinforced with HSSWM-HPM systems. Pham and Hao (2016) reviewed the performance of concrete structures strengthened with FRP systems subjected to impact loads. Their study was an overview of the structural strength of FRP-reinforced concrete beams, slabs, columns and masonry walls. They also evaluated the material properties of FRP under dynamic loading conditions. The outcomes of their work indicated that using FRP can increase load capacity and energy absorption of RC structures. Moreover, the tensile behavior of FRP can increase the strain rate. The experimental study did clearly show the effect of dynamic loads on the debonding mechanism or the FRP rapture strain. Furthermore, many studies have done in this field like Banthia and Mindess (2012). They have investigated the behavior of RC beams under quasi-static and impact loading conditions. They performed experiments at the University of British Columbia. They tested 12 samples of reinforced concrete beams which two of them were under quasi-static loading and others were under impact loadings. Also, they strengthened one beam in quasi-static and impact loading with GFRP sheets. The result showed that the load capacity of beam under quasi-static is higher than beams subjected to dynamic loading. Watstein (1953) performed dynamic tests on reinforced concrete beams, the results showed the compressive strength of concrete increase 85 to 100% under dynamic loads in comparison to that the staics conditions. Khalighi (2009) studied the bond between fiber reinforced polymer and concrete under Quasi-Static and impact loadings. They performed experimental tests on FRP reinforced concrete beams and indicated an increase in the bearing capacity of the beams.

## Model Development

The following sections illustrate the process used to develop the FEM model to analyze retrofitted beams subjected to impact load conditions.

### *Finite Element Model*

The ABAQUS software implementation for modeling of RC beams subjected to impact loading conditions follows the basic model developed by Soleimani *et al.* (2007; Soleimani, 2007). In this model, two types of loading conditions were considered including quasi-static and impact loads. The ABAQUS model uses 3D 8-node linear isoperimetric elements with reduced integration. The hammer is modeled by a solid element with its rigid property applied as Rigid Body interaction. In this case, a Reference Point (RP) is considered at the center of the hammer in which whole elements are rigid to the point. Moreover, the loading conditions are applied as displacement-control at the reference point. The model was validated against 1 m long beam (0.8 m span). Details of the beam are shown in Fig. 1. It is simply supported beam and loaded by a point load at the center (Fig. 1). The longitudinal, transvers bars and mechanical properties of the beam are tabulated in Table 1. The values of $f_y$, $f_u$ and $f_{ys}$, $f_{us}$, M-10 and ϕ4.75 are also shown in the table, respectively.

Moreover, loading conditions and configurations of the FRP bars used in the modeling are shown in Table 2. This table has two sets of data; one is BS (Quasi-Static) data and the second one is impact (as BI-height of hammer). Rate of impact was controlled by the velocity of the drop hammer which was controlled by the drop height of the hammer. All beams were reinforced with CFRP and GFRP bars.

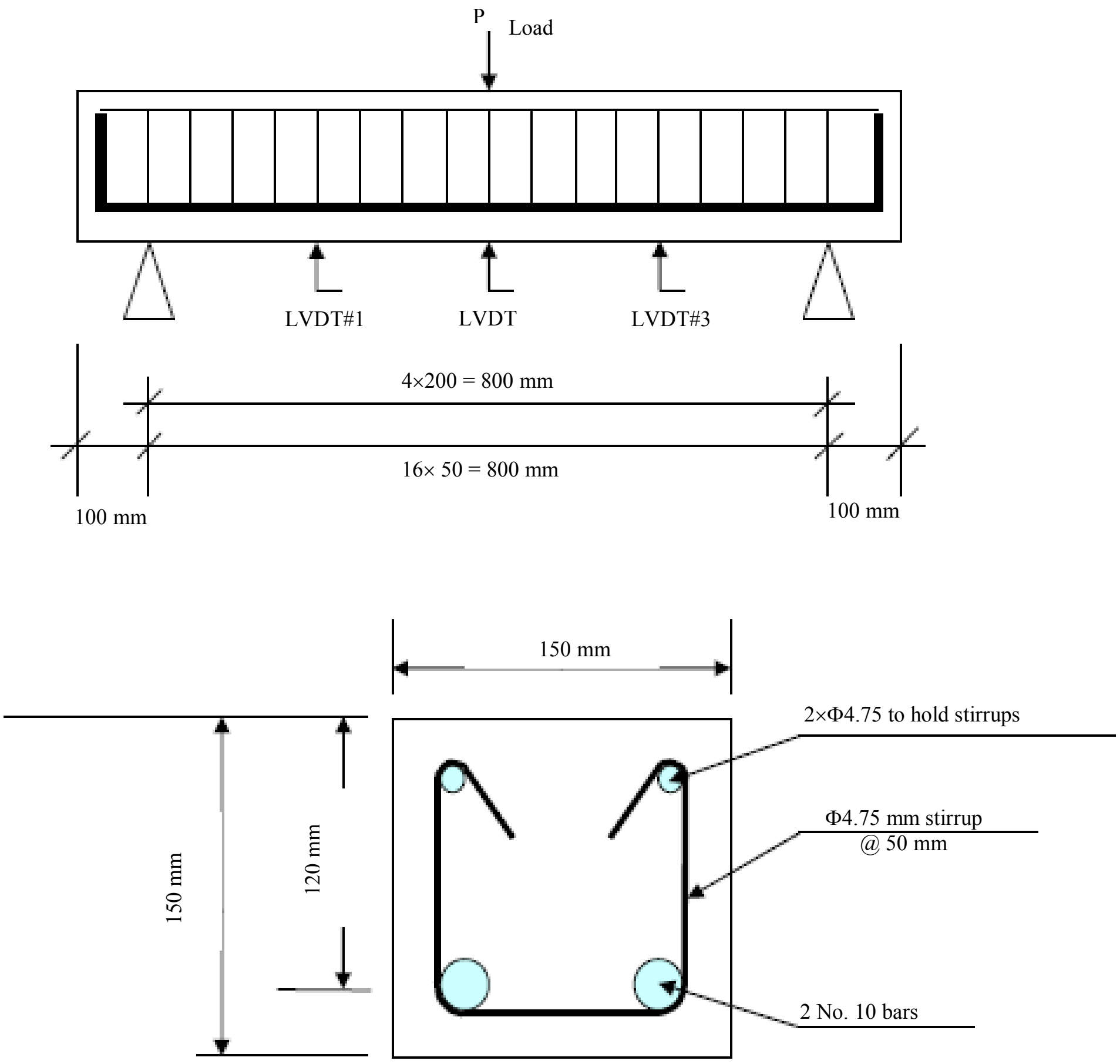

**Fig. 1:** Section details of RC beams

**Table 1:** RC beams properties (Soleimani, 2007; Soleimani *et al.*, 2007)

| Parameter | Definition | Value | Unit |
|---|---|---|---|
| b | Width of compression face of member | 150 | mm |
| H | Overall depth of beam Distance from extreme | 150 | mm |
| D | compression fiber to centeoid of tension reinforcement | 120 | mm |
| $f'_c$ | Specified compressive strength of concrete | 44 | MPa |
| $f_y$ | Specified yield strength of tension reinforcement | 474 | MPa |
| $f_{ys}$ | Specified yield strength of shear reinforcement | 600 | Mpa |
| $f_u$ | Specified ultimate strength of tension reinforcement | 720 | MPa |
| $f_{us}$ | Specified ultimate strength of shear reinforcement | 622 | MPa |
| ϕ4.75 | Area of reinforcement | 18.1 | |
| $A_s$ M-10 | (M-10 for tension and 4.75 for shear) | 100 | mm$^2$ |

**Table 2:** Loading and reinforcing condition properties for FEM software (Soleimani *et al.*, 2007)

| Beanm number | Quasi-static loading | Impact loading drop height, h (mm) 400 | 500 | 600 | 1000 | 2000 | Velocity (m/s) | GFRP bars | CFRP bars |
|---|---|---|---|---|---|---|---|---|---|
| BS | ☑ | - | - | - | - | - | - | ☑ | ☑ |
| BI-400 | - | ☑ | - | - | - | - | 2.80 | ☑ | ☑ |
| BI-500 | - | - | ☑ | - | - | - | 3.13 | ☑ | ☑ |
| BI-600 | - | - | - | ☑ | - | - | 3.43 | ☑ | ☑ |
| BI-1000 | - | - | - | - | ☑ | - | 4.43 | ☑ | ☑ |
| BI-2000 | - | - | - | - | - | ☑ | 6.26 | ☑ | ☑ |

**Table 3:** Specifications of rebar used in accordance with regulations (ACI, 2006)

| Bars type | Density (N/m$^3$) | Tensile strength (MPa) | Module of elasticity (GPa) | Yield strain % | Rupture strain % |
|---|---|---|---|---|---|
| CFRP | 150-160 | 600-3690 | 120-580 | NA | 0.5-1.7 |
| GFRP | 125-210 | 483-1600 | 35-51 | NA | 1.2-3.1 |

The mechanical properties of the CFRP and GFRP bars are shown at Table 3.

### *Concrete Stress-Strain Model*

The inputs of ABAQUS require known geometry and mechanical properties of materials, especially for concrete material. Concrete parameters are usually based on empirical equations that relate stress to its corresponding strains. In this study, the concepts of smeared crack and concrete damage plasticity models (Jankowiak and Tlodygowski, 2005; Voyiadjis and Abu-Lebdeh, 1994; Abu-Lebdeh and Voyiadjis, 1993) were used to relate stresses to stains. These models were used due to their versatile usefulness in different types of loading conditions such as: static, dynamic or monotonic and cyclic loadings. The models considered compressive and tensile stress-strain under its damage states.

For ABAQUS Model, Fig. 2 is adopted to define the post failure stress-strain relationship of concrete. The input parameters were Young's modulus ($E_0$), stress ($\sigma_t$), cracking strain $\left(\tilde{\varepsilon}_t^{ck}\right)$ and the damage parameter values ($d_t$) for the relevant grade of concrete. The cracking strain $\left(\tilde{\varepsilon}_t^{ck}\right)$ can be calculated by Equation (1):

$$\tilde{\varepsilon}_t^{ck} = \varepsilon_t - \varepsilon_{0t}^{el} \tag{1}$$

where, $\varepsilon_{0t}^{el} = \sigma_t / E_0$ the elastic-strain corresponding to the undamaged material, $\varepsilon_t$ is total tensile strain.

Moreover, the plastic strain $\left(\tilde{\varepsilon}_t^{pl}\right)$ for tensile behavior of concrete can be defined as shown in Equation 2:

$$\tilde{\varepsilon}_t^{pl} = \tilde{\varepsilon}_t^{ck} - \frac{d_t}{1-d_t}\frac{\sigma_t}{E_0} \tag{2}$$

A typical diagram for compressive stress-strain relationship with damage properties is illustrated in Fig. 3. The inputs are stresses ($\sigma_c$), inelastic strains $\left(\tilde{\varepsilon}_c^{in}\right)$ corresponds to stress values and damage properties ($d_c$) with inelastic in tabular format. It should be noted that the total strain values should be converted to the inelastic strains using Equation (3):

$$\tilde{\varepsilon}_c^{in} = \varepsilon_c - \varepsilon_{oc}^{el} \tag{3}$$

For the compressive behavior of concrete, the elastic strain $\varepsilon_{oc}^{el} = \sigma_c / E_0$ where $\varepsilon_{oc}^{el}$ corresponds to the strain of undamaged material and $\varepsilon_c$ is the total compression strain. In addition, the plastic strain values $\left(\tilde{\varepsilon}_c^{pl}\right)$ is calculated using Equation (4):

$$\tilde{\varepsilon}_c^{pl} = \tilde{\varepsilon}_c^{in} - \frac{d_c}{1-d_c}\frac{\sigma_c}{E_0} \tag{4}$$

### *MATLAB Strain Incorporation*

In MATLAB section, we continue the work of Roudsari *et al.* (2018) who performed some theoretical evaluations on the compressive and tensile behavior of concrete. In their study, the ultimate stress and its corresponding strain were used as input for MATLAB. They were determined either from experimental tests or from theoretical formulas. Furthermore, the compression and tension diagram were utilized to generate data needed to optimize strain rate at an increment of 0.0001. The bottom line here is that, using the formula and coding in MATLAB give the compression stress values that correspond with its strain rate and it will be continued to the ultimate strain. This process had been done in tensile behavior of the concrete, too. On the other hand, the ABAQUS software's input is only plastic part of diagrams, so according the ACI standard, the linear and nonlinear parts were separated at 45% of maximum compression strength (Roudsari *et al.*, 2018).

### *Post-Failure Stress-Strain Relation*

In ABAQUS software, the post-failure behavior of reinforced concrete member can be approximated using the relation shown in Fig. 4. It is worth mentioning that, in sections with little or no reinforcing elements, the meshing plays an important role due to the sensitivity of the results to the mesh which can possibly have negative or positive effects on the outputs. As such, using an appropriate mesh can display cracks more accurately and more visibly.

The interaction between the reinforcing bars and the surrounding concrete induce stresses may generate more tensile stress on the concrete elements. In this study, stiffening is introduced in the cracking model to simulate this interfacial interaction. It is completely depending on reinforcement density, relative size of the concrete aggregate to rebar diameter, quality of the bond between the rebar and the concrete and the type of mesh. In normal concrete, the strain at failure is typically 10 4 in/in, however, tension stiffening can reduce the stress to a total strain of about 10 3 (Hillerborg *et al.*, 1976).

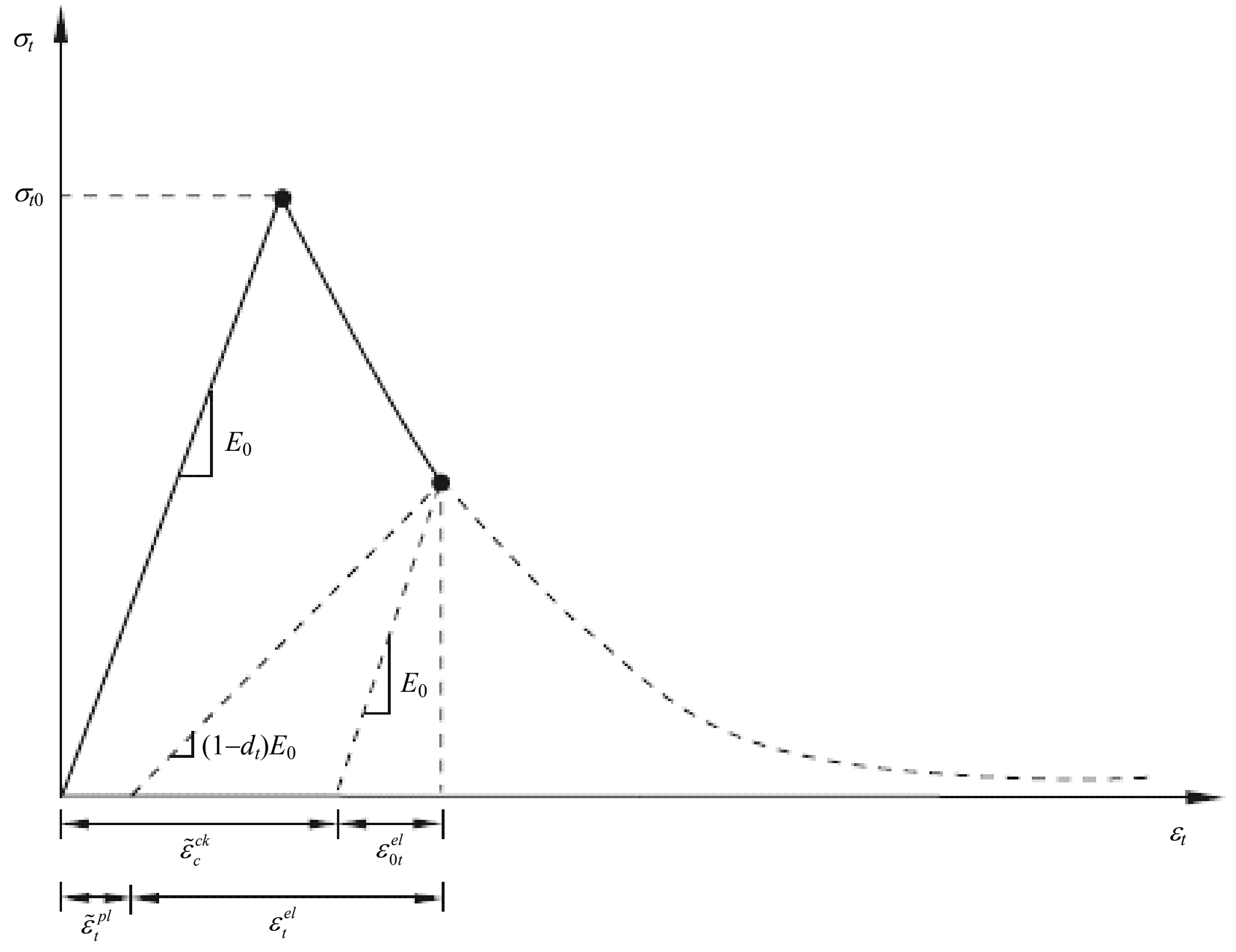

**Fig. 2:** Tension stiffening parameters (Jankowiak and Tlodygowski, 2005)

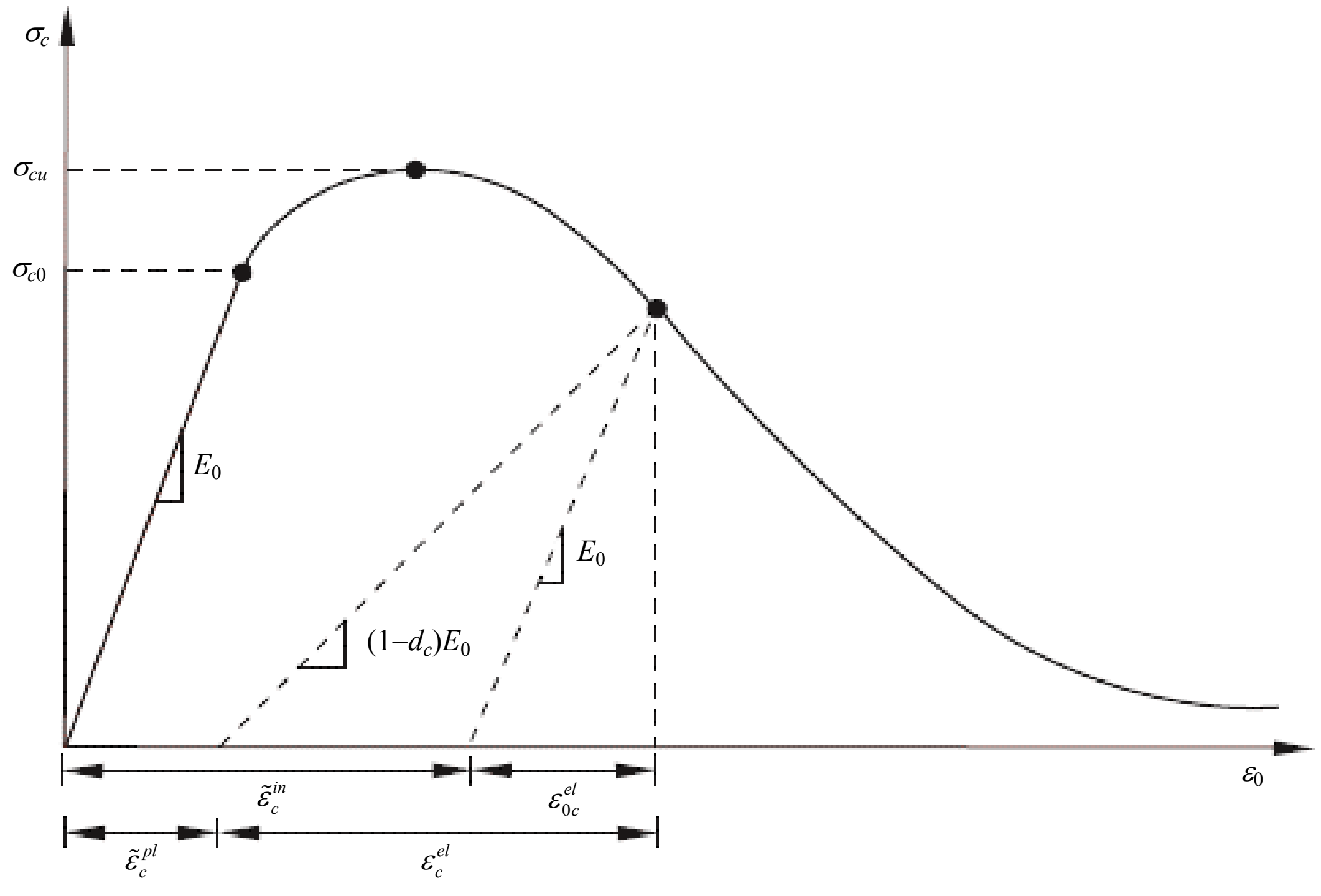

**Fig. 3:** Terms for compressive stress-strain relationship (Jankowiak and Tlodygowski, 2005)

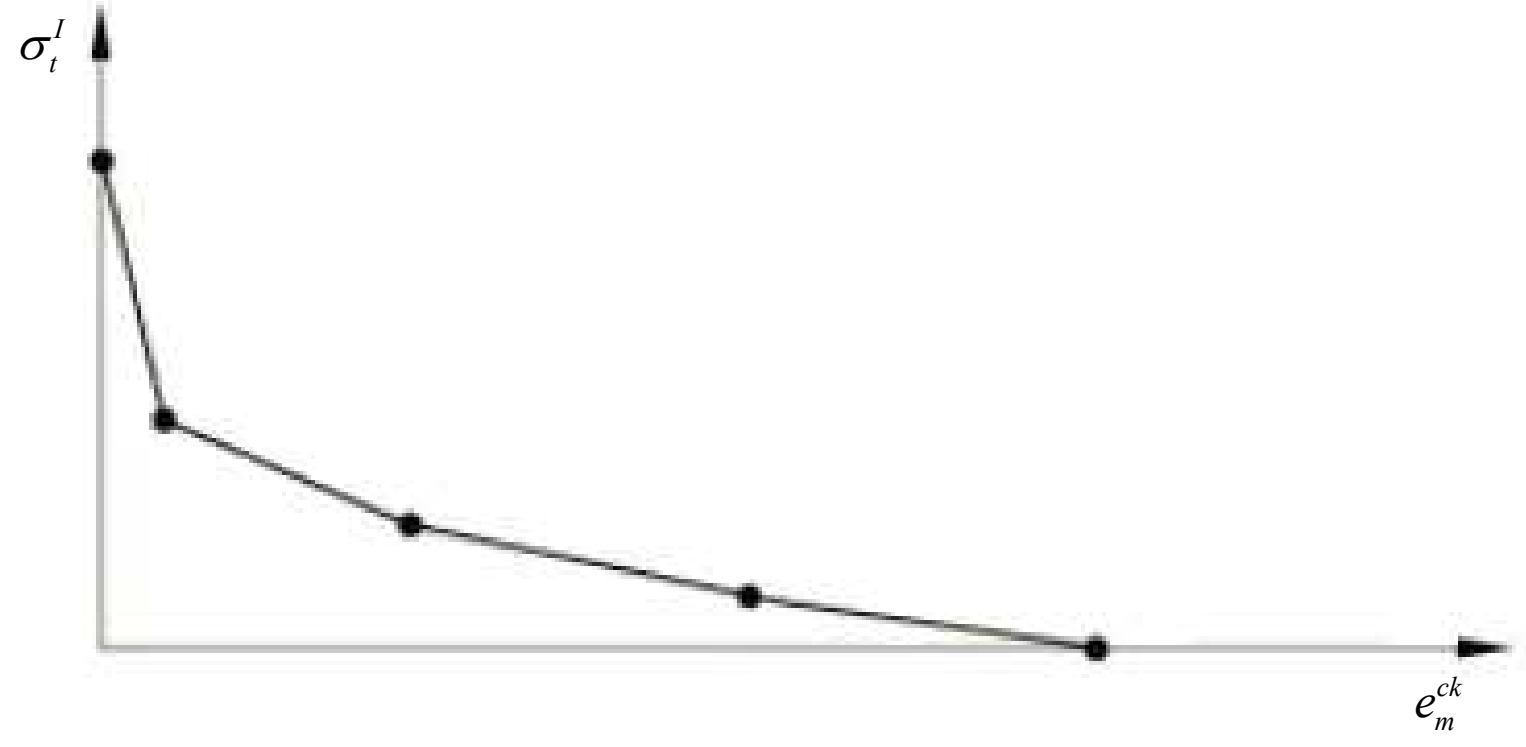

**Fig. 4:** Post-failure stress-strain curve (Hillerborg *et al.*, 1976)

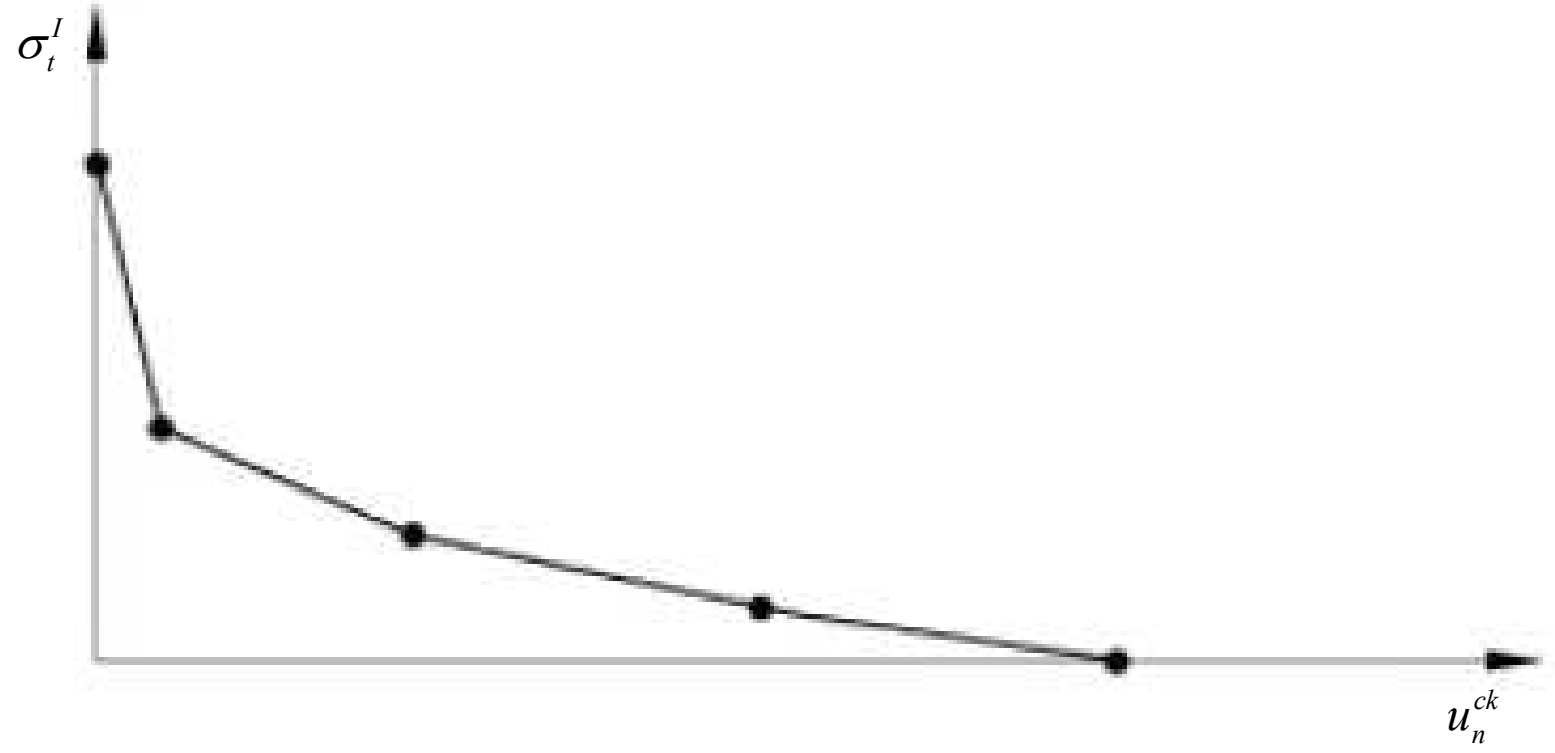

**Fig. 5:** Post failure stress-displacement (Hillerborg *et al.*, 1976)

### *Fracture Energy Cracking Criterion*

In regions where there is no reinforcement, the model uses the same tension stiffening approach described above. This introduces unreasonable mesh sensitivity into the results. However, it is generally accepted that Hillerborg's fracture energy model (Hillerborg *et al.*, 1976) is adequate to allay the concern for different practical purposes. In their model, the energy required to open a unit area of crack in Mode $I\left(G_f^I\right)$ is defined as a material parameter, using brittle fracture concepts. With this approach, the concrete's brittle behavior is characterized by stress displacement response (Fig. 5) rather than stress-strain response. Under tension, a concrete specimen may exhibit small elastic strain cracks across some sections and along its length. This may be determined primarily by the opening at the crack, which does not depend on the specimen's length (Fig. 5). Alternatively, Mode I fracture energy $\left(G_f^I\right)$ can be specified directly as a material property. In this case, the failure stress, $\left(\sigma_{tu}^I\right)$ can be defined as a tabular function of the associated Mode I fracture energy, assuming linear loss of strength after cracking (Fig. 6).

The crack normal displacement at which complete loss of strength takes place is, therefore $U_{no} = \frac{2G_f^I}{\sigma_{tu}^I}$. Typical values of range from 40 N/m for normal concrete (with a compressive strength of approximately 20 MPa, to 120 N/m for concrete (with a compressive strength of approximately 40 MPa.

It should be noted that the $G_f^I$ function is used as a parameter for the concerte's tensile behavior so that it can be determined by ABAQUS documentation. It can be divided into three different categories (Hillerborg *et al.*, 1976): (1) $G_f^I$ = 40 MPa if compressive strength ≤20 MPa; (2) $G_f^I$ = 20 MPa If the compressive strength ≥40 MPa; and (3) for compressive strength between 40 MPa and 120 MPa, then a linear interpolation can be used. Further, the tensile stress is defined as follows:

$$\sigma_{ti} = F_{ct}.\exp\left[\frac{1}{\gamma_t}\left(\frac{F_{ct}}{E}\right) - \varepsilon_i\right] \quad (5)$$

where, $\varepsilon_i$ is the strain rate which is based on number of increments. In fact, for every increment, there is a different value for both strain and stress.

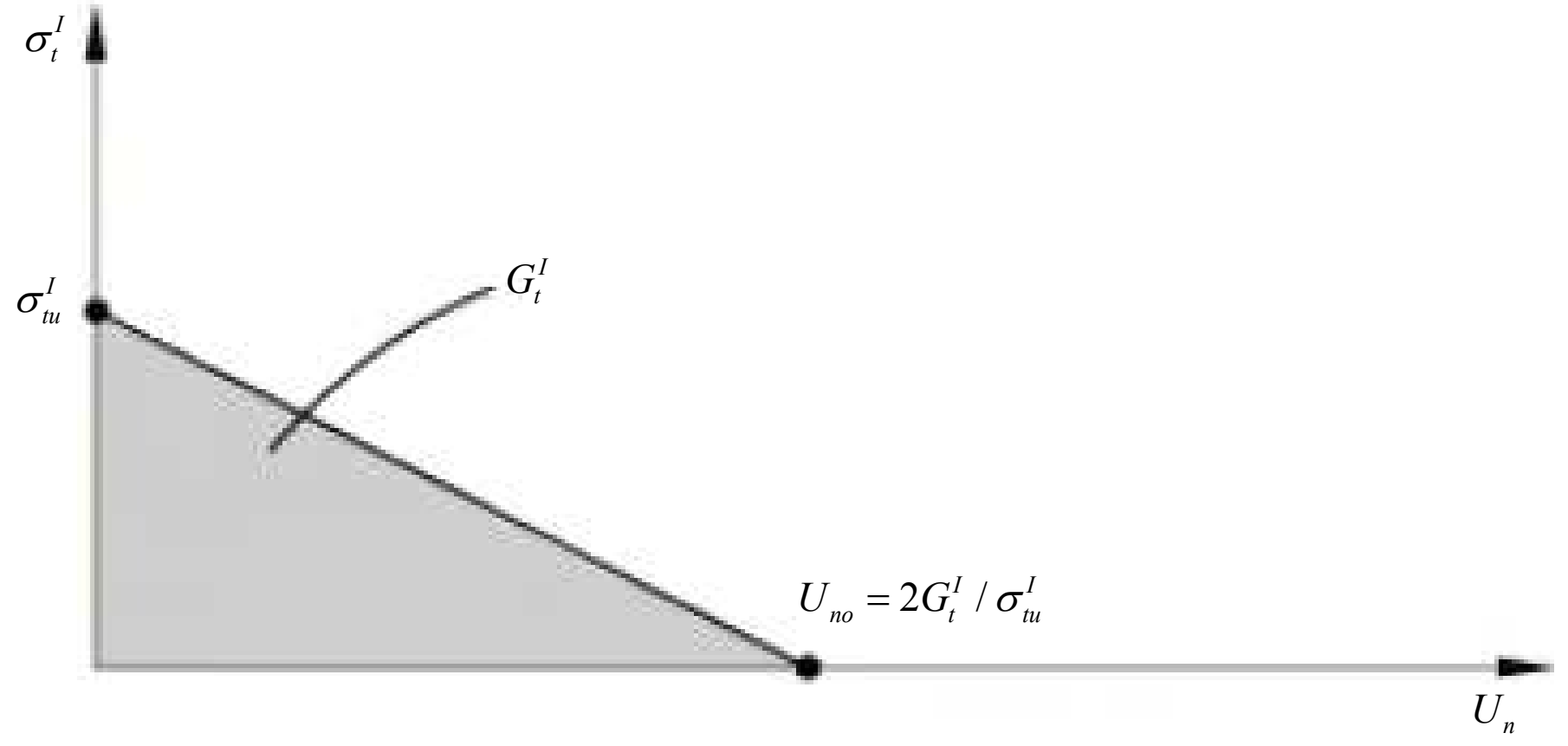

**Fig. 6:** post-failure stress-fracture energy curve (Hillerborg *et al.*, 1976)

The term $\gamma_t$ can be determined using the function:

$$\gamma_t \frac{GFI}{i_{eq} \times F_{ct}} - \frac{F_{ct}}{2E} \tag{6}$$

The damage parameter for the tensile behavior of concrete can been expressed as follows:

$$d_{ti} = 1 - \left( \frac{\varepsilon_{teli}}{\varepsilon_i - \varepsilon_{tpl}} \right) \tag{7}$$

$\varepsilon_{teli}$ is the elastic strain at the corresponding tension stress, it may be defined as:

$$\varepsilon_{teli} = \frac{\sigma_{ti}}{E}, \varepsilon_{tpli} = 146 \times \varepsilon_i^2 + 0.523 \times \varepsilon_i \tag{8}$$

The tensile parameters can now be solved by the above functions and the compressive parameters can also be defined. Ultimately, only plastic parameters are needed as inputs for the ABAQUS software. In the function below, $\varepsilon_i$ the strain incrementation and $\varepsilon_c$ is the strain at the maximum compressive stress:

$$\sigma_{ci} = \frac{E \times \varepsilon_i}{1 + \left( \frac{E}{E_S} - 2 \right) \times \frac{\varepsilon_i}{\varepsilon_c} + \times \left( \frac{\varepsilon_i}{\varepsilon_c} \right)^2} \tag{9}$$

Finally, the function of compression damage $d_{ci}$ can be defined by:

$$d_{ci} = 1 - \frac{\varepsilon_{celi}}{\left( \varepsilon_i - \varepsilon_{cpli} \right)} \tag{10}$$

In this case, $\varepsilon_{celi}$ is the elastic strain which can be defined as: $\varepsilon_{celi} = \frac{\sigma_{ci}}{E}$.

Also, the plastic strain $\varepsilon_{cpli}$ is defined as:

$$\varepsilon_{cpli} = \varepsilon_c \times \left( 0.166 \times \left( \frac{\varepsilon_i}{\varepsilon_c} \right)^2 + 0.132 \times \left( \frac{\varepsilon_i}{\varepsilon_c} \right) \right) \tag{11}$$

It should be noted that these functions are the most important and useful functions in calculating plasticity parameter of concrete damage, but they need to be verified. The work of Jankowiak and Tlodygowski (2005) and the coding program of Roudsari *et al.* (2018) were used in this study for verification. In their numerical study, they obtained stress-strain curves where the maximum strength and its corresponding strain were 50 MPa and 0.0122, respectively (Fig. 7). As shown, the difference between the two graphs is insignificant and thus it may be concluded that the parameters are correct. At this step, the linear segment of the diagram should be separated from the nonlinear part. This is because the plastic output is needed for inputting in ABAQUS. Therefore, as it has been noted that the segment up to 45% of the compressive strength represents the linear portion; the second part has to be modified so that all compressive strengths and their corresponding strains will move to the initial coordinate (0, 0). The outputs of MATLAB for ABAQUS software are shown in Fig. 8 and 9.

### *ABAQUS Modeling*

Three dimensional models with eight nodes by reduced integration (C3D8R) was used for modeling of concrete. Also, truss elements (T3D2) were used for creating longitudinal and transvers FRP reinforcements. The concrete damage plasticity model was used for concrete behavior and a nonlinear model was used for FRP bars. Because of brittle failure of FRP bar, in addition to modulus of elasticity, only ultimate stress and its correspond stain were used since there is no yield stress in the diagram. In other word for making two linear diagrams of FRP bar in ABAQUS, the yield stress is considered a little bit lower than ultimate stress. The interaction between the concrete and bars is modeled by the embedded region. Also, in order to avoid the scattering result, a Reference Point (RP) is defined at the center of each support. Moreover, the coupling is assigned the RP to sum output from whole nodes of bottom surface of the support (Nicoletto and Riva, 2004).

## *Loading Conditions*

The model considers two groups of loading conditions. The first group is quasi-static loadings defined in term of Dynamic-Implicit and the second group is the impact loadings defined as Dynamic-Explicit. For quasi-static case, the loading hammer was located at the top center of the beam and displacement was computed by defining a node (defined a set in ABAQUS) at the bottom center of the beam. Also, the hammer used for impact loading on the middle of beam with different velocity and height. Both hammers for quasi-static and impact loading were considered to be solid and rigid bodies. Moreover, the loading for both conditions were assigned on the top of hammer by defining a load-displacement control parameter and corresponding loading rate. This was modelled by inputting a tabular amplitude which started from zero and continued by 80% of loading value in 0.7 sec to reach 100% of total load in one second. Moreover, the velocity of impact loading is assigned by Velocity/Angular Velocity in ABAQUS. It should be noted that Reference Point (RP) is defined for all loadings and support's reactions. The bottom supports are hinge which the degree of freedom of U1, U2 and U3 has considered zero and the ends of beam are pinned in order to avoid rotation of beam. In order to avoid rotation of beam for impact loading, two steel yokes are considered exactly parallel and same location of bottom hinge supports. The interaction of bars and concrete and boundary condition have shown for quasi-static and impact loading at Fig. 10.

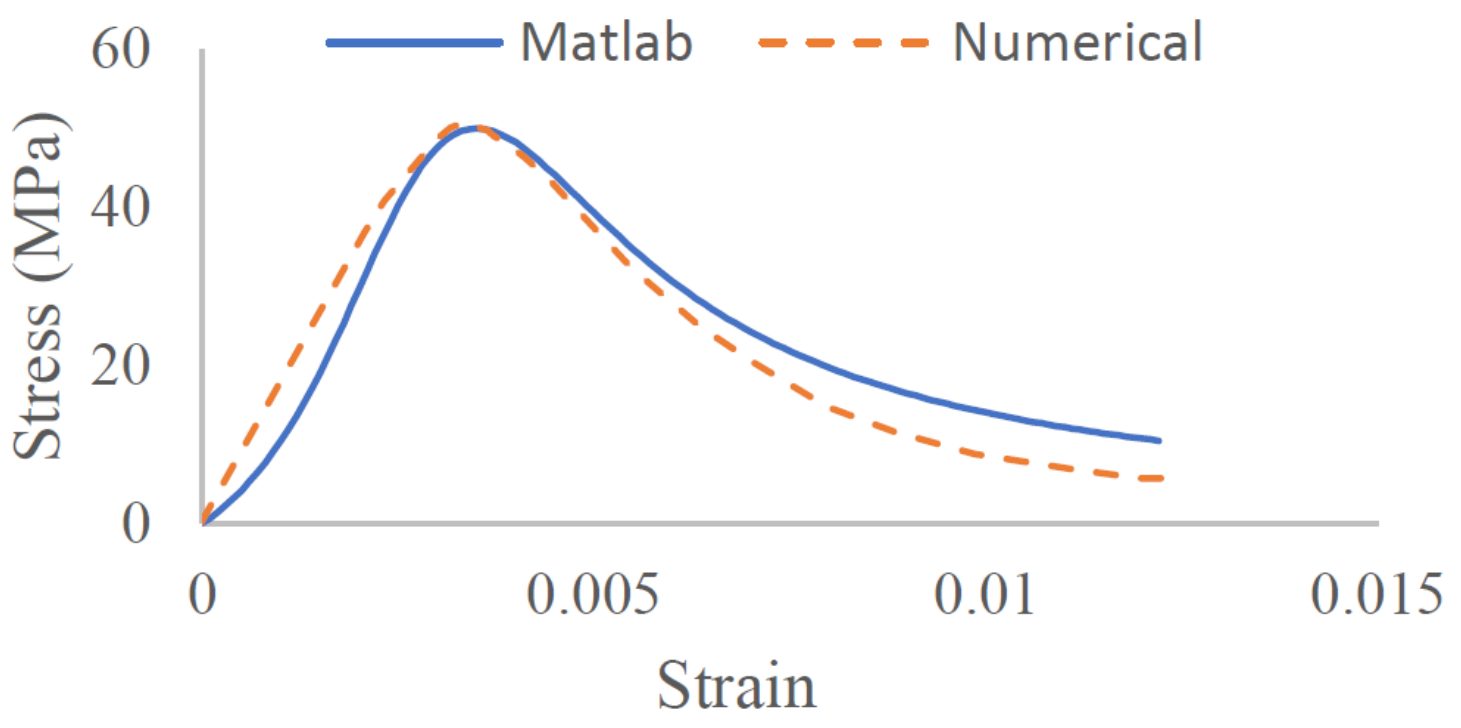

**Fig. 7:** Compressive strain-stress – FEM and experimental models (Roudsari *et al*., 2018)

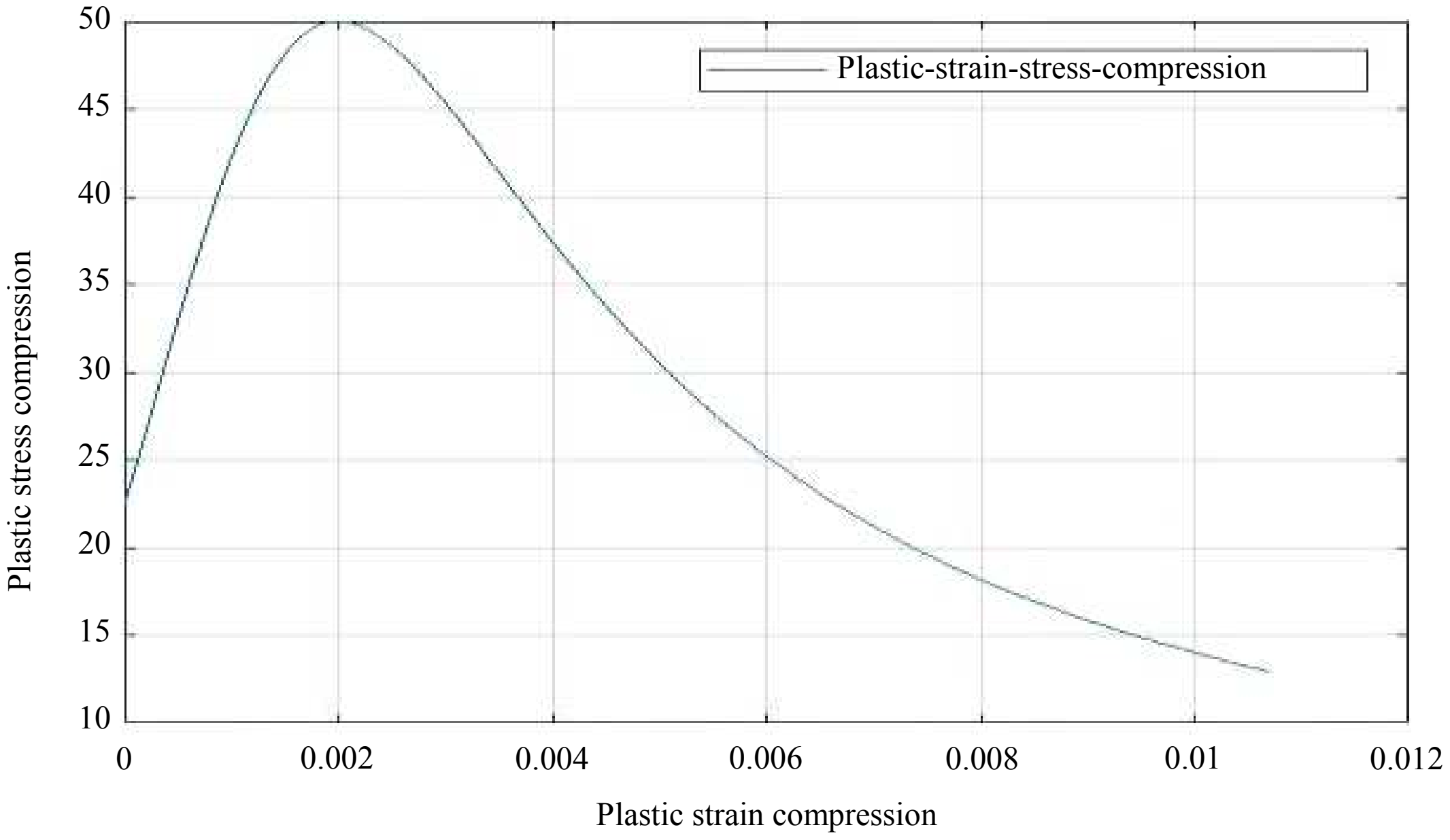

**Fig. 8:** Output of MATLAB for ABAQUS (Roudsari *et al*., 2018)

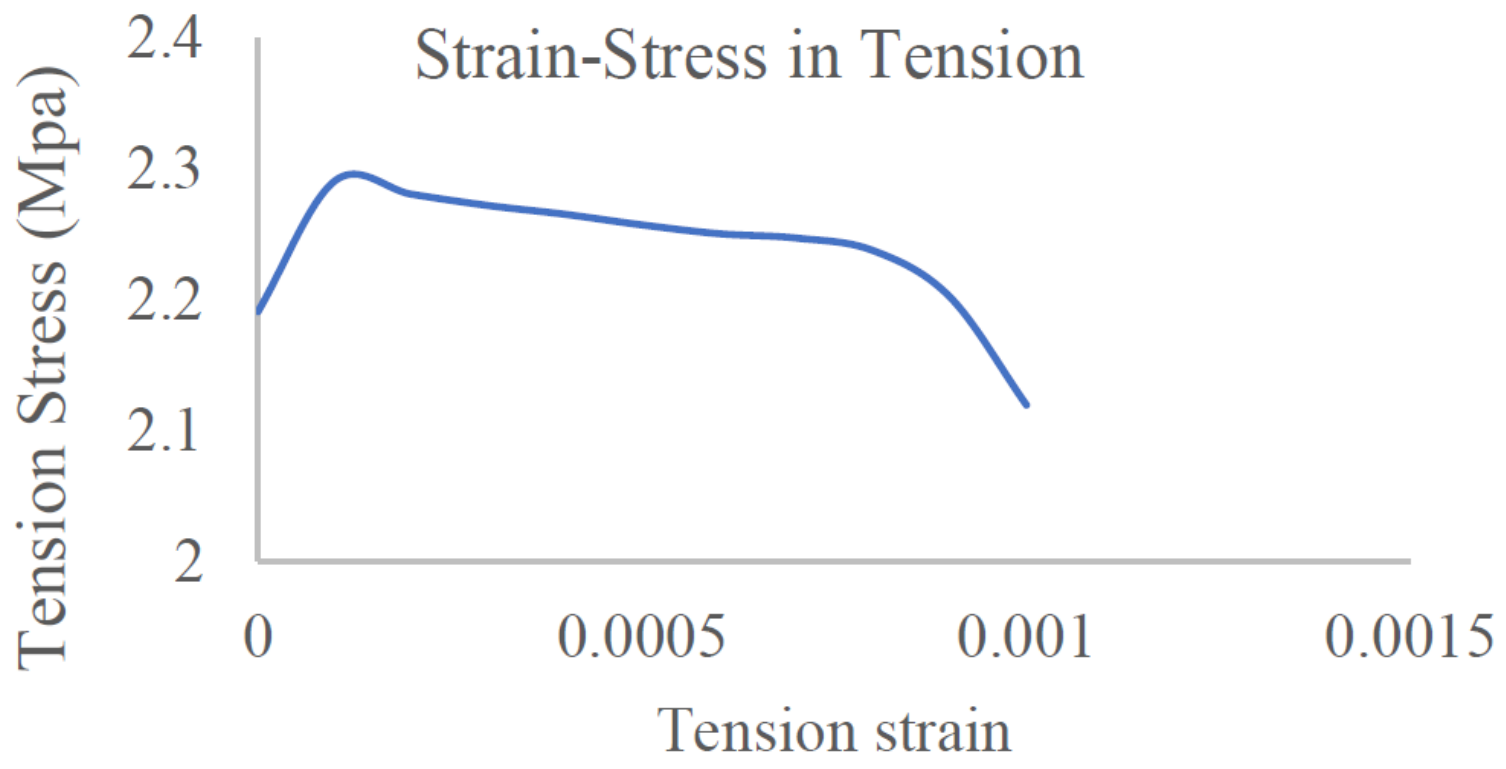

**Fig. 9:** Tension stress-strain diagram by MATLAB

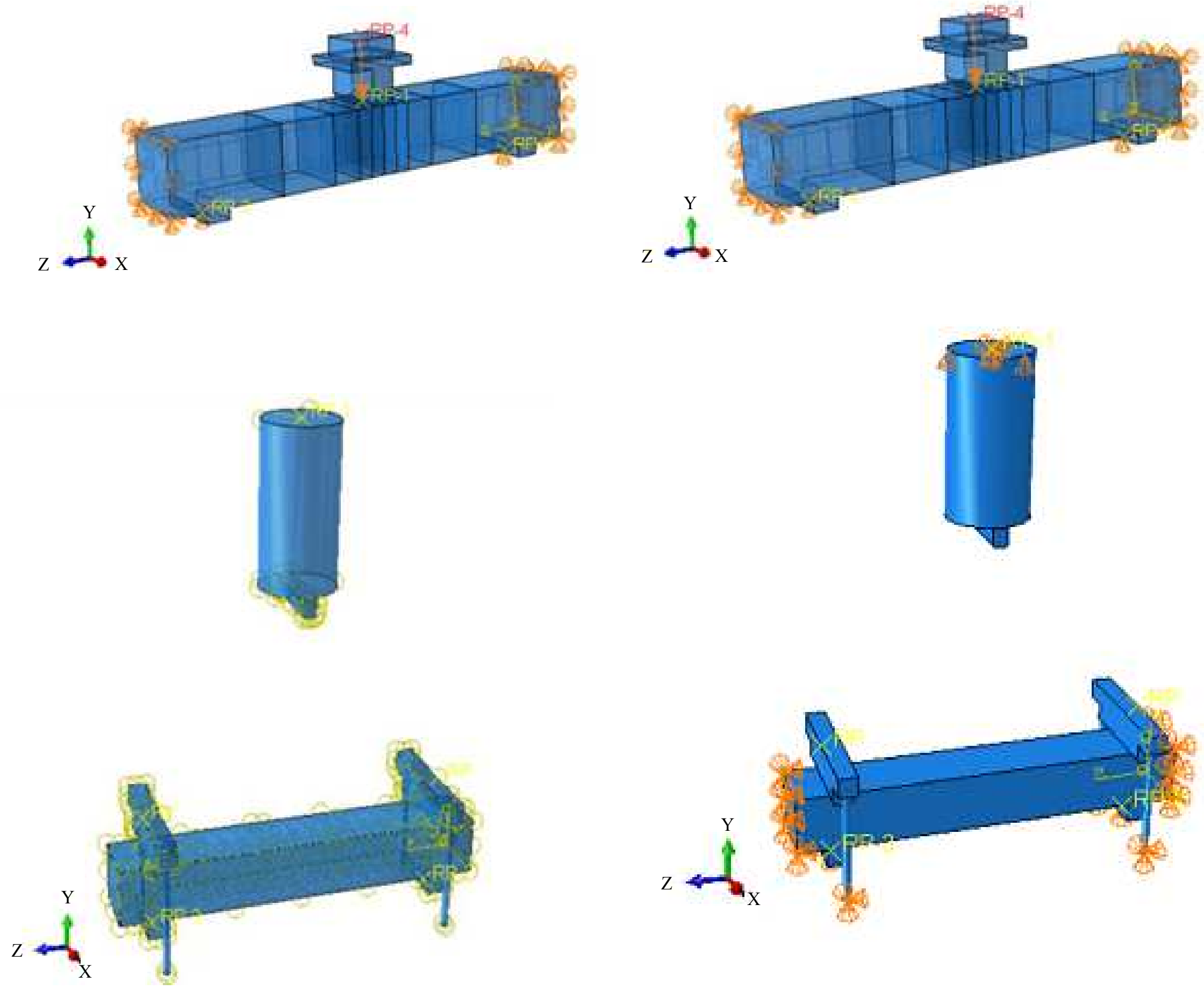

**Fig. 10:** Details of modeling in ABAQUS

### *Output of FRP Bars Modeling in ABAQUS*

In this section, results of the FEM modeling are shown in Fig. 11-16. These figures display the load displacement diagram of FPR reinforced concrete beams under quasi-static loading and impact loading.

### *Model Verification*

For model verifications, the authors use two different types of experiments. The first experimental work was generated from Soleimani's thesis which is regarding concrete beams reinforced with steel bars and retrofitted by GFRP sheets, while the second verification was generated from Goldston *et al*. (2016) experimental test.

### *Verification with Steel bars and GFRP Sheets*

In this section, the authors validated ABAQUS results with the experimental tests. The impact and Quasi-Static loading parameters were the same. Properties of steel bars and GFRP sheets are shown in Table 4.

The loading conditions of impact and quasi-static loading in laboratory are shown in Fig. 17. GFRP is used for retrofitting in term of flexural and shear behavior. The width of layout is 1.5 meters and length of 0.75 meters and its thickness is 0.353 millimeters. U wrapped is used for controlling of shear behavior in three faces of beam. Mechanical and physical properties of GFRP is shown in Table 5. Furthermore, the mechanical properties of steel are: Module of elasticity 200 GPa, tensile strength 483 to 690 MPa and its rupture strain 6-12%, respectively. It is necessary to declared that Hashin Damage is used to define parameters and lamina is used to define modules of elasticity and shear modules in different directions.

**Table 4:** Loading condition and reinforcing properties of experimental tests (Soleimani, 2007)

| Name of beam | Quasi-static loading | Impact loading drop height, h (mm) 400 | 500 | 600 | 1000 | 2000 | Velocity (m/s) | GFRP sheets | Steel bars |
|---|---|---|---|---|---|---|---|---|---|
| BS | ☑ | - | - | - | - | - | - | - | ☑ |
| BS-GFRP (Sheet) | ☑ | - | - | - | - | - | - | ☑ | ☑ |
| BI-400 | - | ☑ | - | - | - | - | 2.80 | - | ☑ |
| BI-500 | - | - | ☑ | - | - | - | 3.13 | - | ☑ |
| BI-600 | - | - | - | ☑ | - | - | 3.43 | - | ☑ |
| BI-600-GFRP (Sheet) | - | - | - | ☑ | - | - | 3.43 | ☑ | ☑ |
| BI-1000 | - | - | - | - | ☑ | - | 4.43 | - | ☑ |
| BI-2000 | - | - | - | - | - | ☑ | 6.26 | - | ☑ |

**Table 5:** GFRP Properties on the basis of Hashin (Hillerborg *et al*., 1976)

| Tensile strength in fiber direction (Mpa) | Compressive strength in fiber direction (Mpa) | Tensile strength perpendicular to the fiber (Mpa) | Compressive strength perpendicular to the fiber (Mpa) | Longitudinal shear strength (Mpa) | Transverse shear strength (Mpa) |
|---|---|---|---|---|---|
| 3660 | 2803 | 240 | 426 | 89.7 | 89.7 |

**Table 6:** Comparison between the base shear and displacement numerical and laboratory samples

| Difference displacement, FEM Vs. experiments (%) | Difference base shear forces, FEM Vs. experiments (%) | Specimen |
|---|---|---|
| 1.25 | 0.06 | BS |
| 4.5 | 20.00 | BI-400 |
| 1.8 | 3.2.0 | BI-500 |
| 4.6 | 6.15 | BI-600 |
| 4.7 | 3.7.0 | BI-1000 |
| 2.75 | 0.3.0 | BI-2000 |
| 1.4 | 0.5.0 | BS-GFRP |
| 4.35 | 19.35 | GFRP |

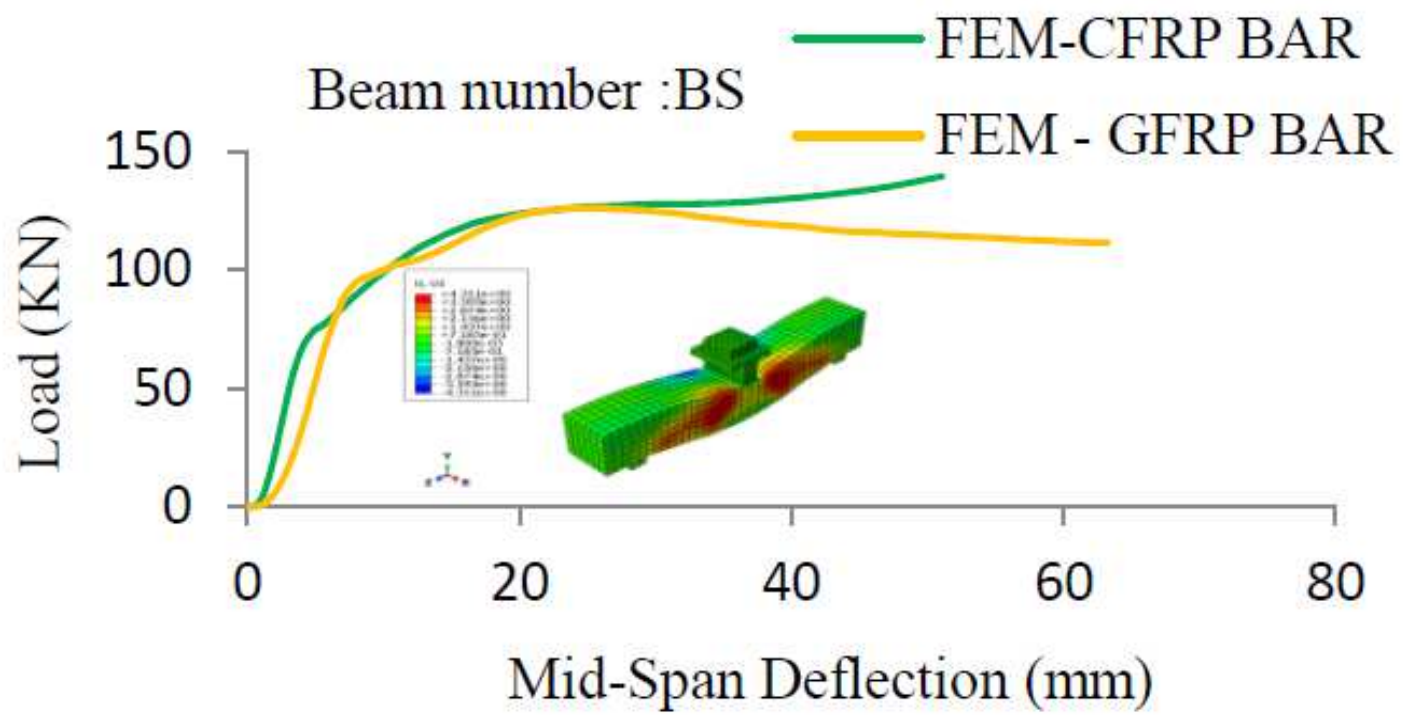

**Fig. 11:** Load-displacement diagram for BS and reinforced with carbon and glass rebar

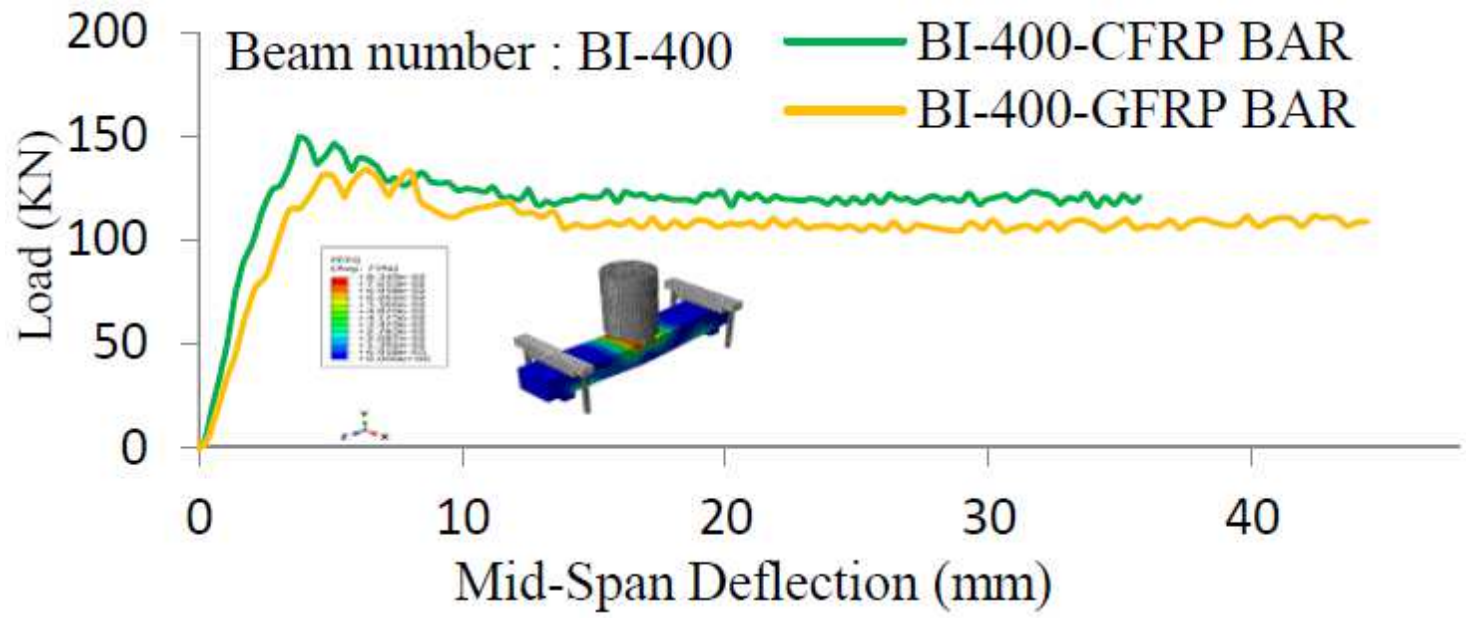

**Fig. 12:** Load-displacement diagram for BI-400 and reinforced with carbon and glass rebar

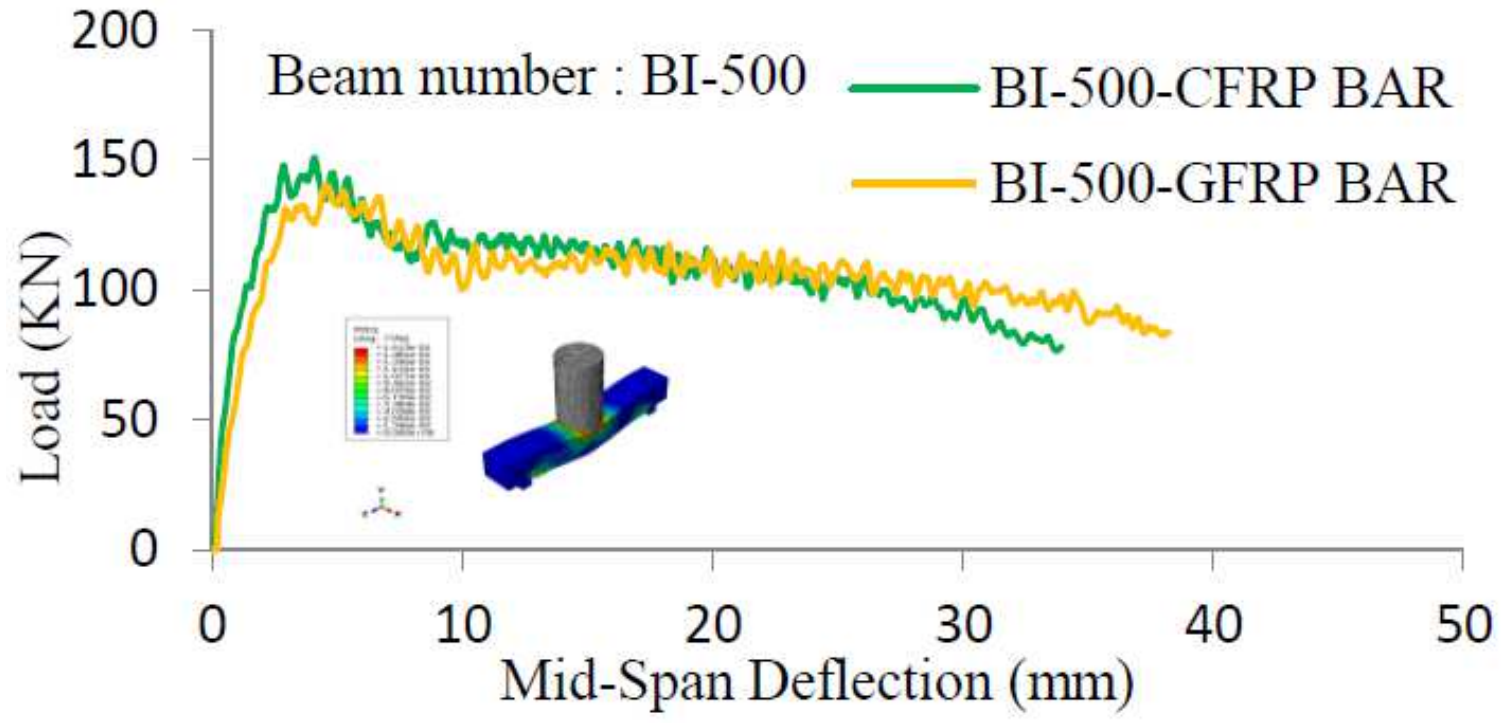

**Fig. 13:** Load-displacement diagram for BI-500 and reinforced with carbon and glass rebar

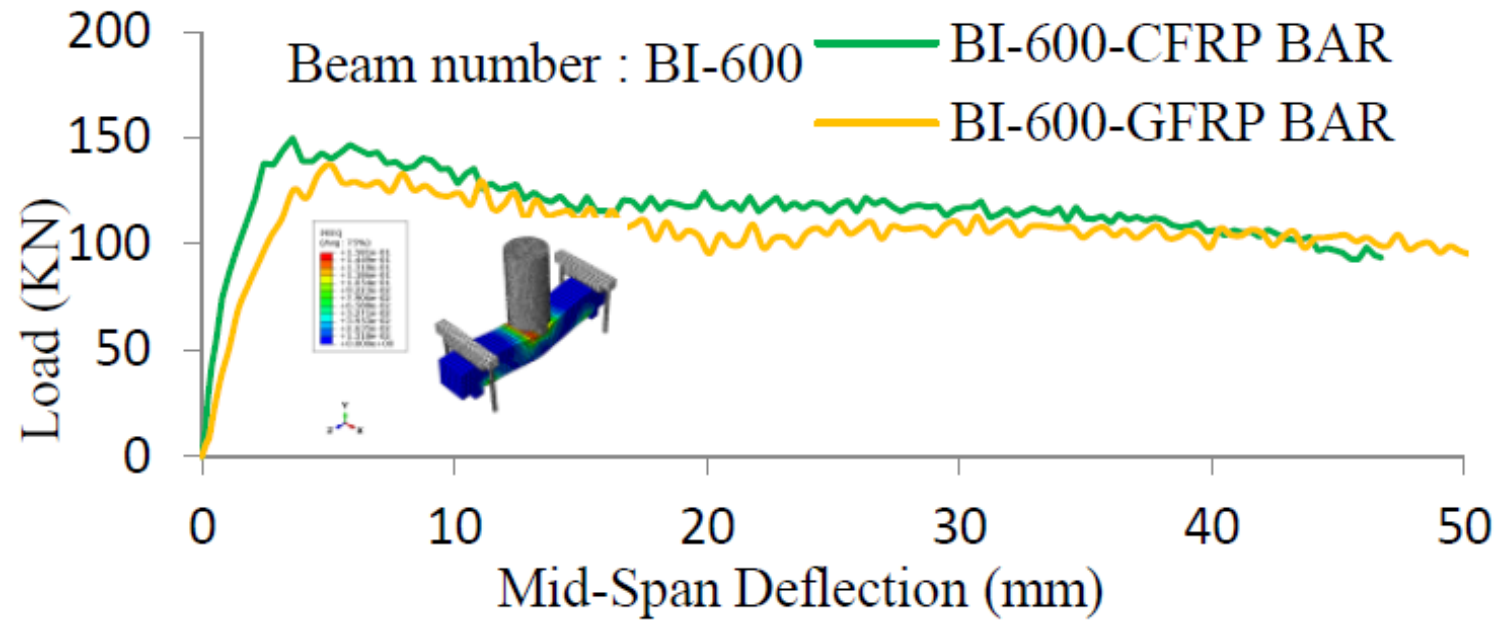

**Fig. 14:** Load-displacement diagram for BI-600 and reinforced with carbon and glass rebar

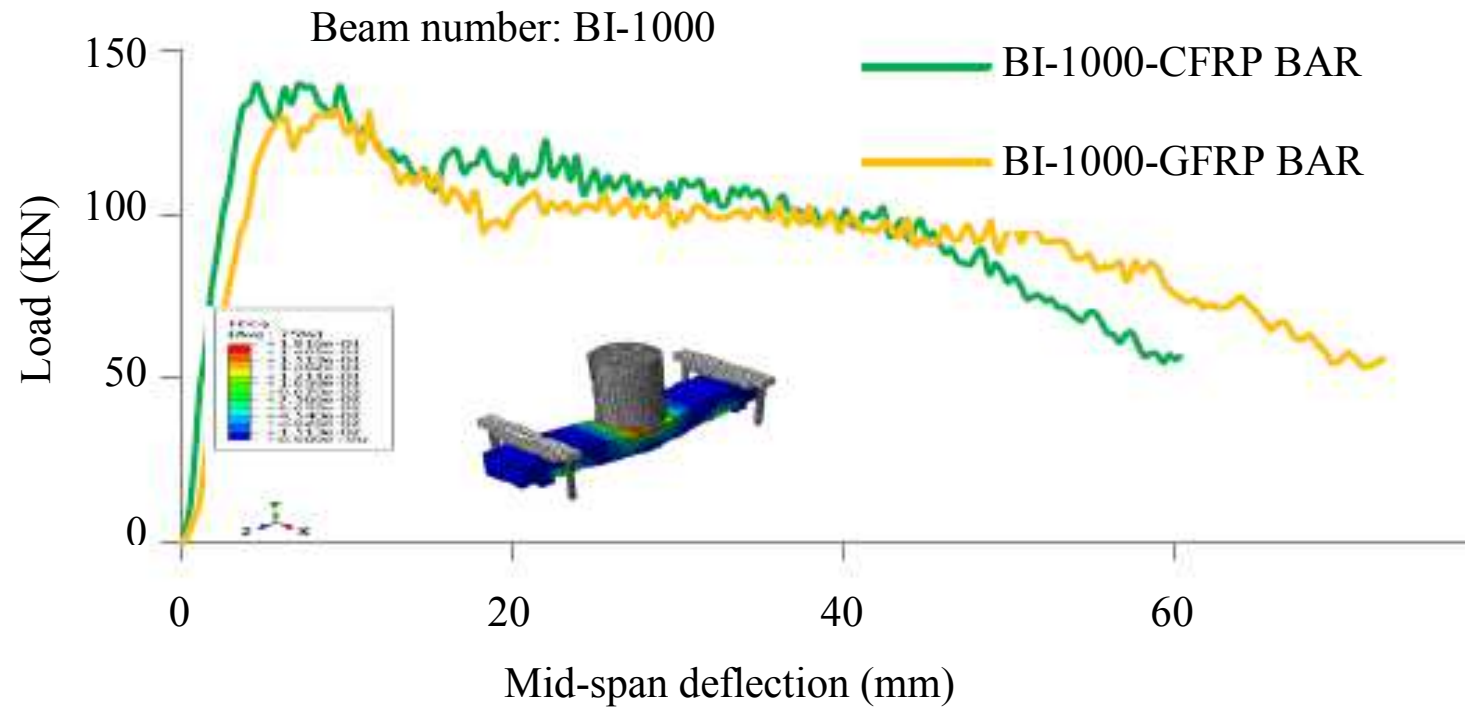

**Fig. 15:** Load-displacement diagram for BI-1000 and reinforced with carbon and glass rebar

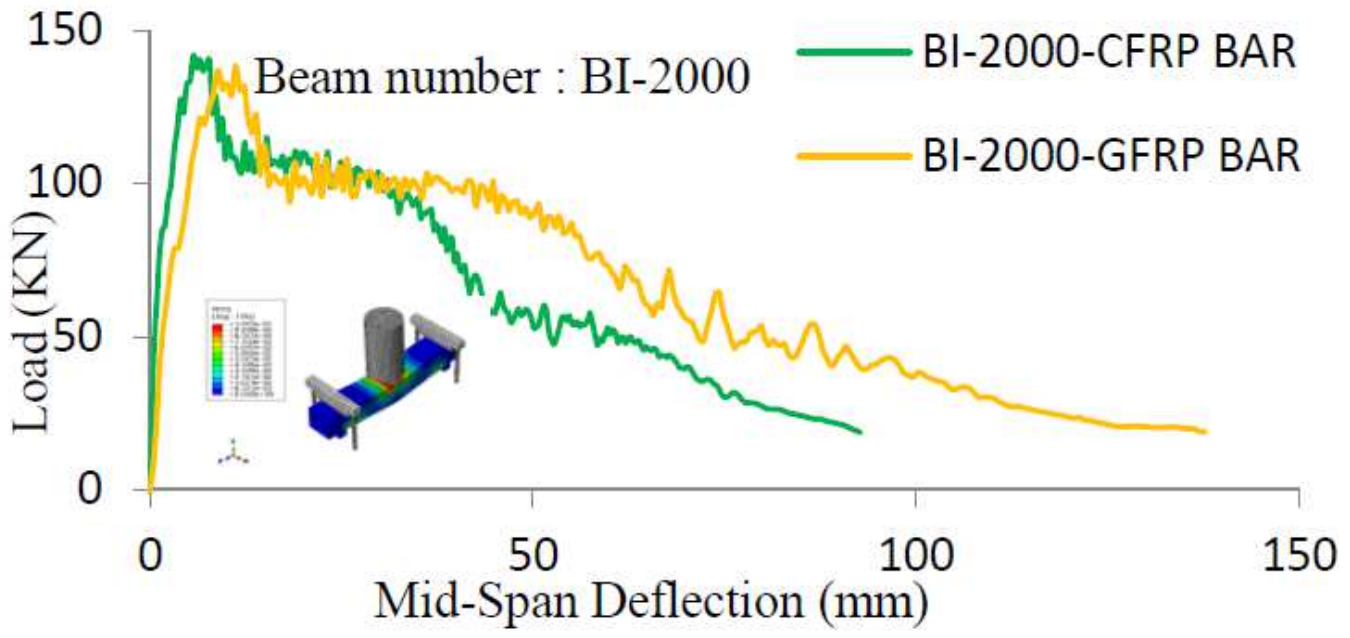

**Fig. 16:** Load-displacement diagram for BI-2000 and reinforced with carbon and glass rebar

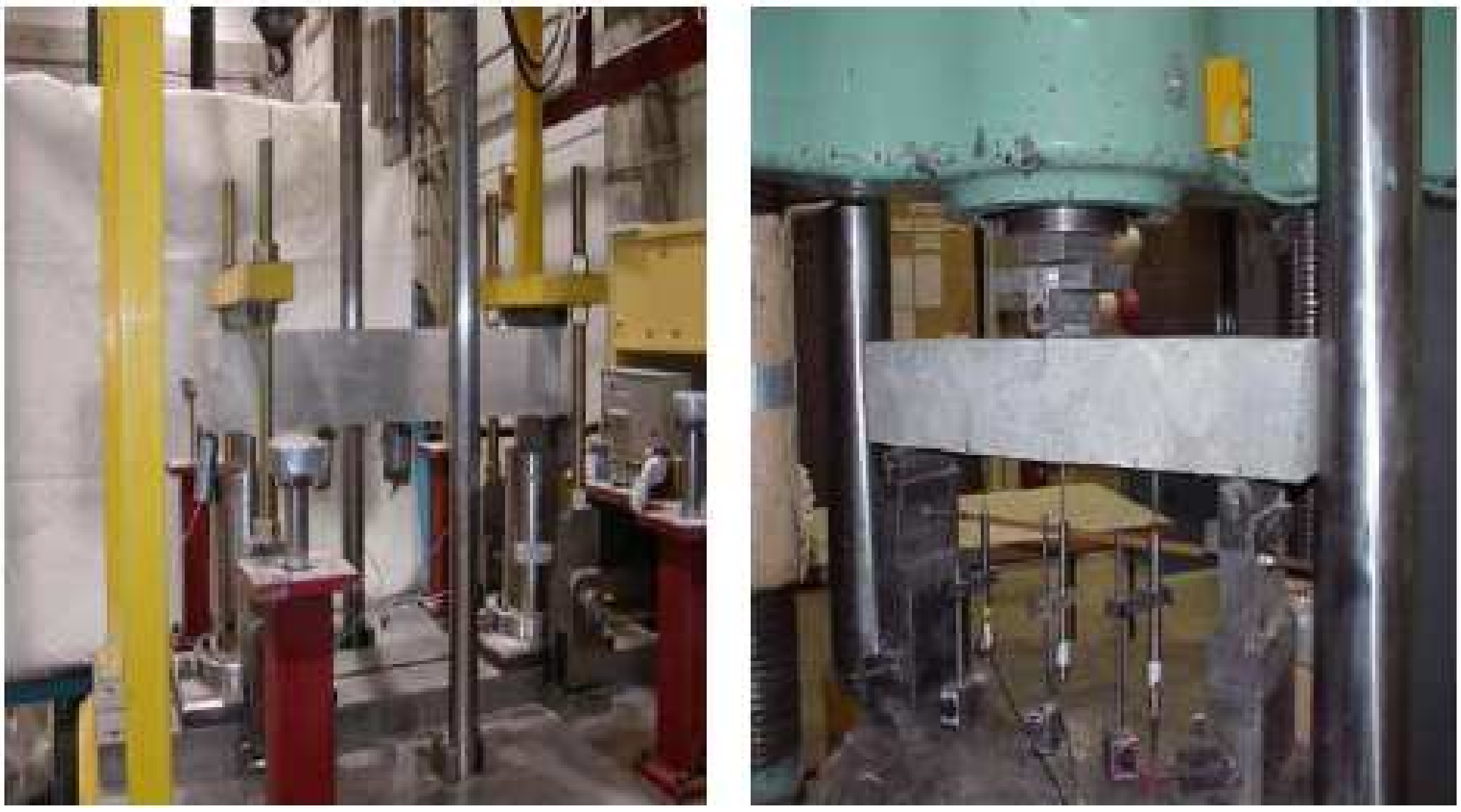

(a) (b)

**Fig. 17:** (a) Quasi-static loading, (b) impact loading condition (Soleimani, 2007)

To verify the model, comparison between ABAQUS modeling and the experimental tests of Soleimani is shown in Fig. 18-25. Also, as shown in Table 6, the difference between finite element modeling and experimental outputs are closely intertwined so that in the case of BS (quasi-static) the maximum difference of base shear in software vs laboratory is about 0.05% and its displacement's differences is less than 1.3%. Also, there is an appropriate difference in results of the impact loading. Results are tabulated in Table 6. As an example, the difference between displacement and base shear for software output and laboratory for BI-2000 is 2.75 and 0.3%, respectively, while these differences are about 1.8 and 3.2% for BI-500.

### *Verification of Concrete Beam Reinforced by GFRP Bar*

Goldston *et al*. (2016) conducted experimental programs which were divided into two different groups, the first group consisted of 6 beams subjected to static loading and second group was under impact loading.

As it can be seen in Fig. 26, three different bars include 6.35 mm (#2), 9.53 mm (#3) and 12.7 mm (#4) were used and generally two GFRP bars located at the top and two others at the bottom of beam. Also, the diameter of steel stirrups is 4 mm at 100 mm were used. The ultimate stress of #2, #3 and #4 (6.35, 9.53, 12.7 mm) bars were 732 Mpa, 1801 Mpa and 1642 Mpa respectively. The moduli of elasticity were 37.5, 53.7 and 47.9 GPa, respectively. The compressive strength of concrete was 40 MPa and its corresponding strain was 0.003. Furthermore, loading was done by spherical ball which was at the center of beam and at the 667 mm of each support and midpoint deflection was calculated by linear potentiometer which was attached at the bottom and center of beam. The loading condition is shown in Fig. 27.

The above specimen's detailing is used to model the GFRP reinforce concrete beam in ABAQUS. As illustrated in Fig. 28, the modeling is done by defining materials and assigning boundary conditions and interactions. it is necessary to mention that the experimental sample with #4 GFRP bars was used to verify the model.

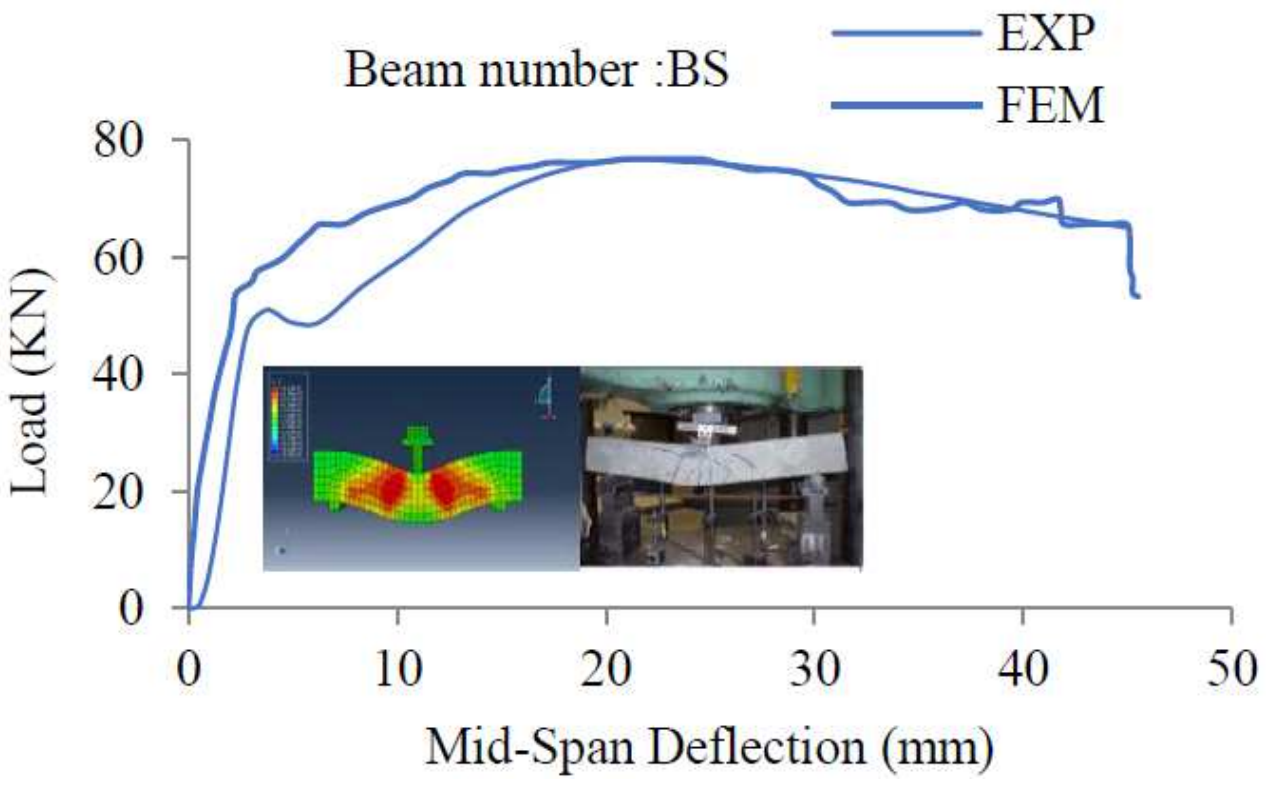

**Fig. 18:** Force-displacement at the ends of beam series

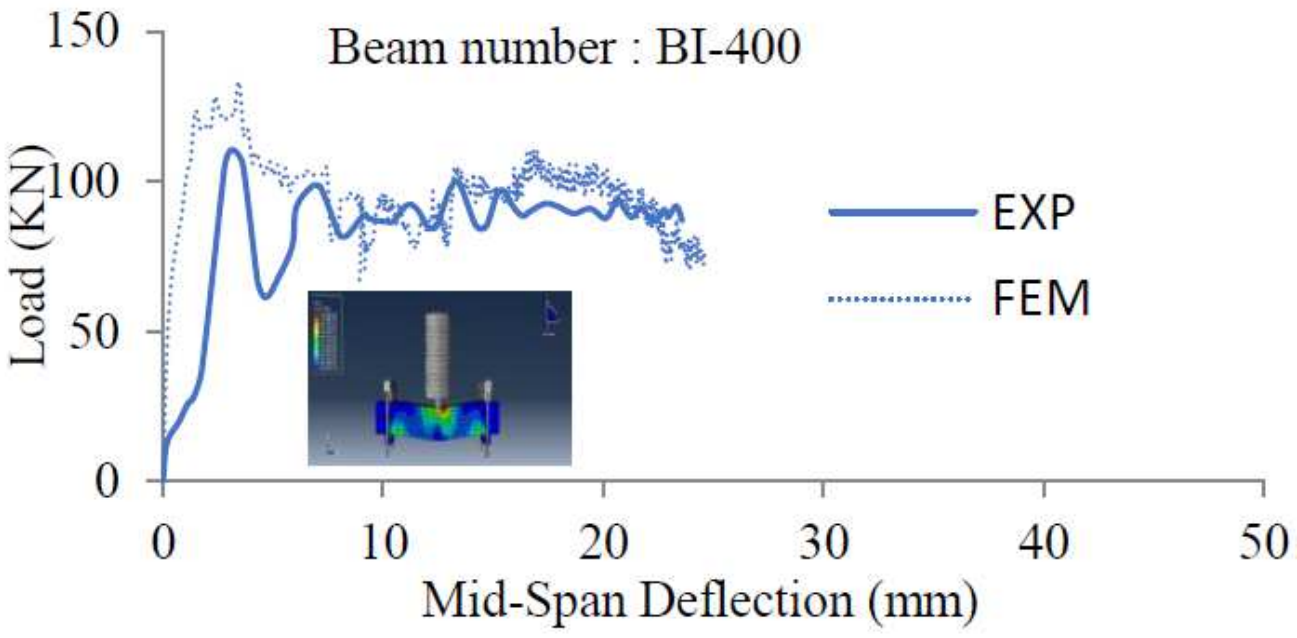

**Fig. 19:** Force-displacement at the ends of beam series BI-400

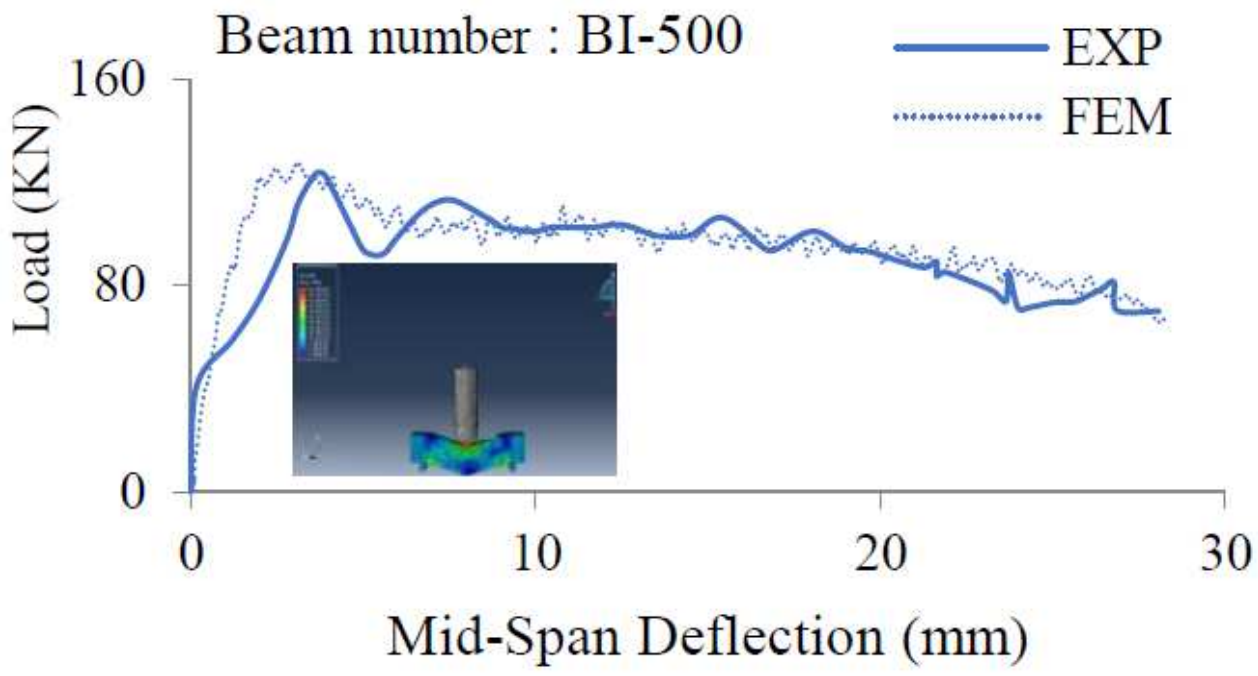

**Fig. 20:** Force-displacement at the ends of beam series BI-500

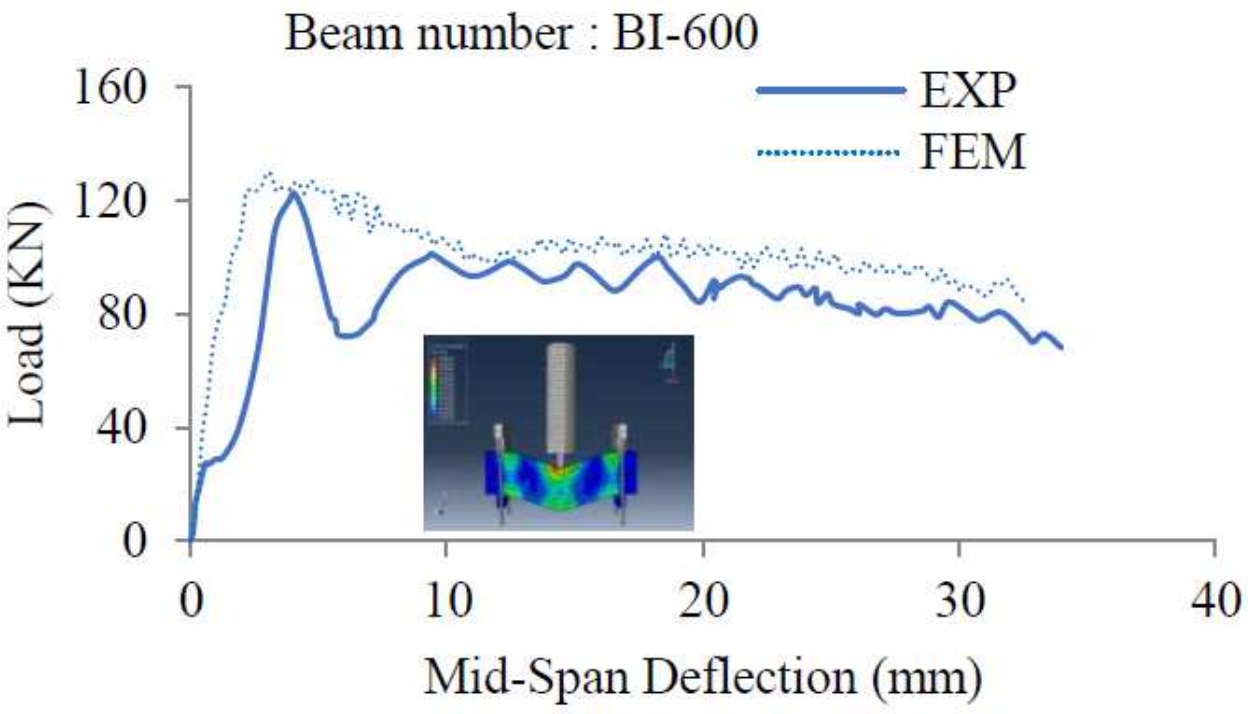

**Fig. 21:** Force-displacement at the ends of beam series BI-600

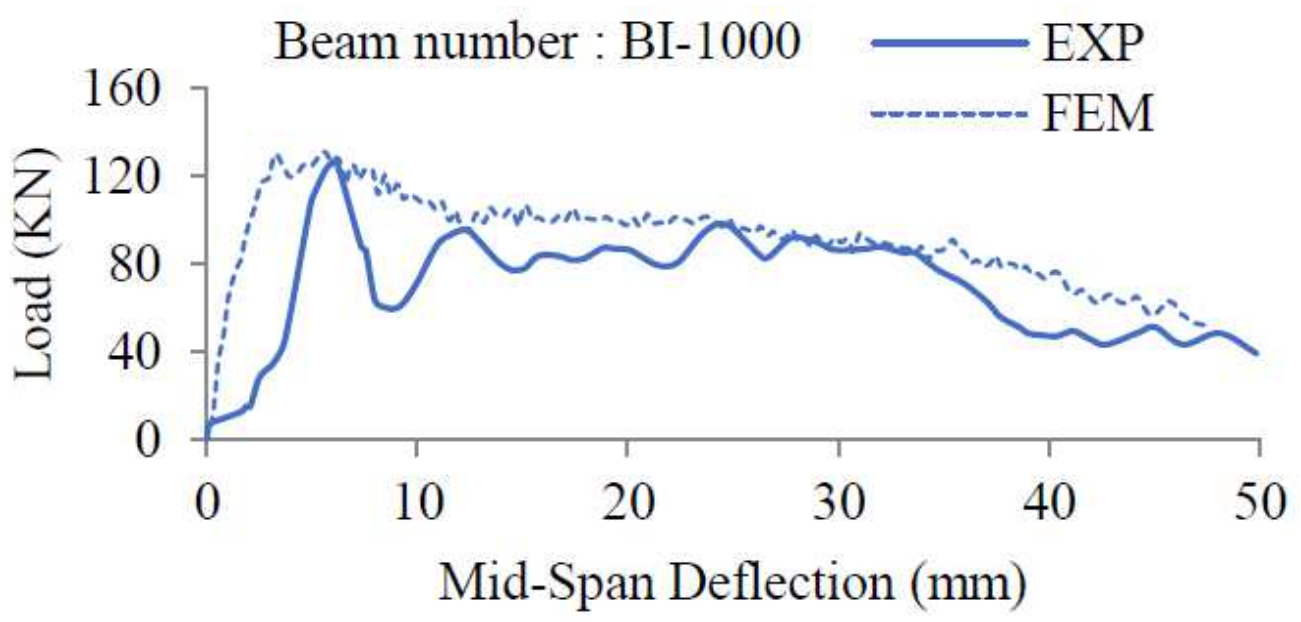

**Fig. 22:** Force-displacement at the ends of beam series BI-1000

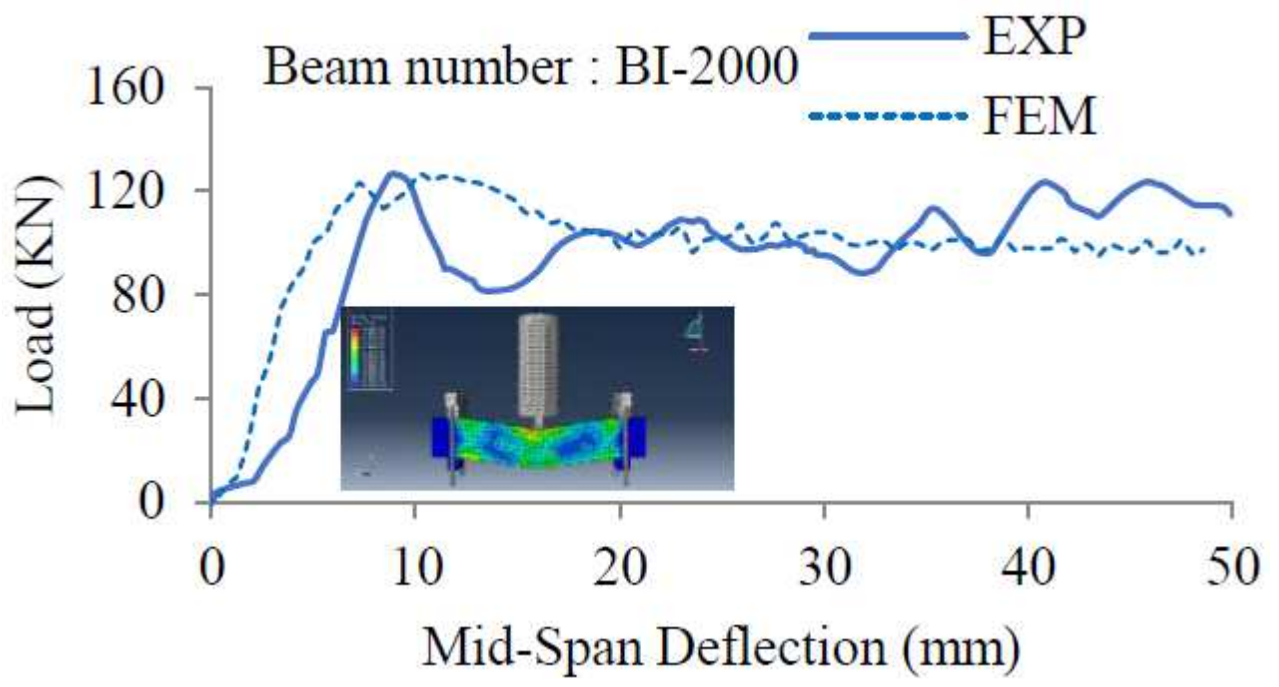

**Fig. 23:** Force-displacement at the ends of beam series BI-2000

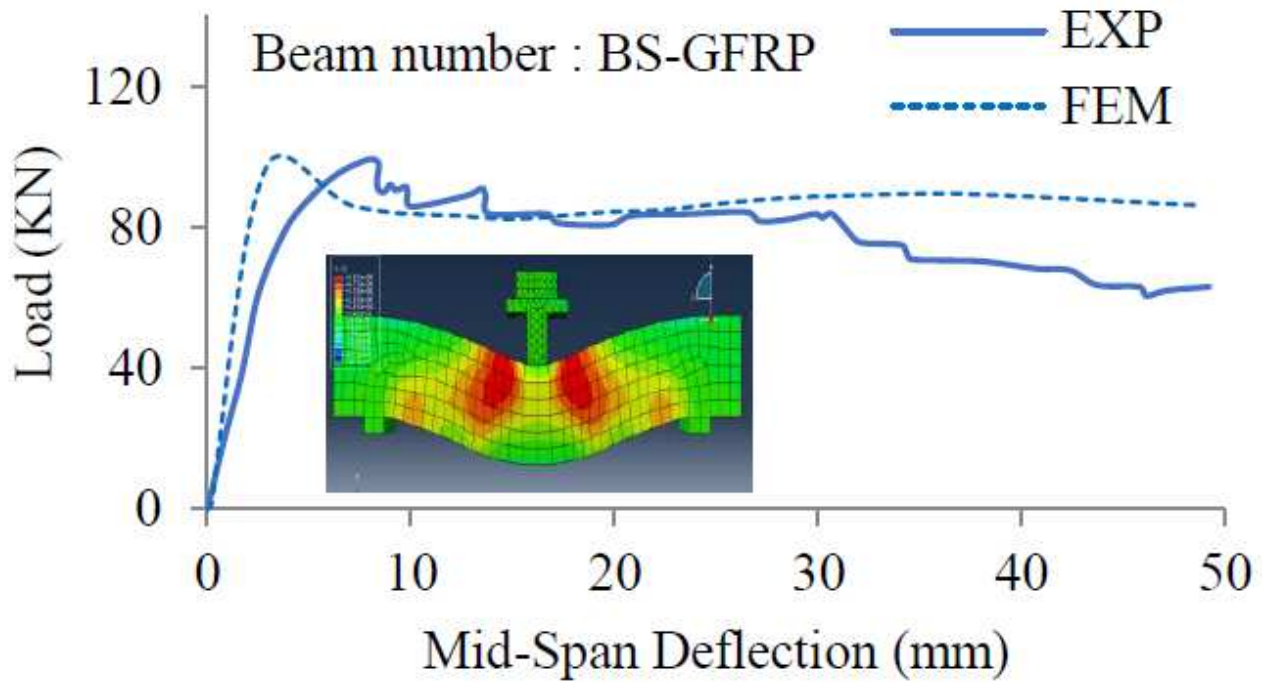

**Fig. 24:** Force-displacement at the ends of beam series BS-GFRP

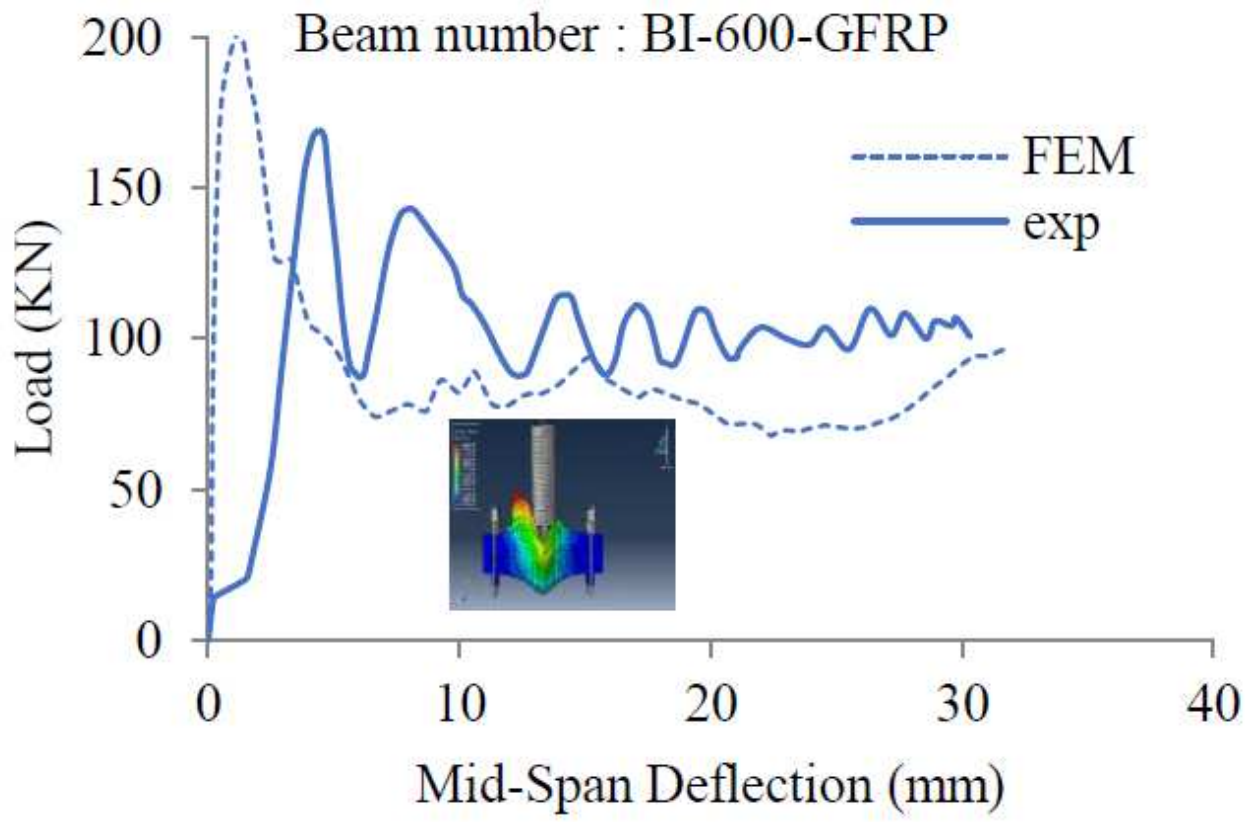

**Fig. 25:** Force-displacement at the ends of beam series BI-600-GFRP

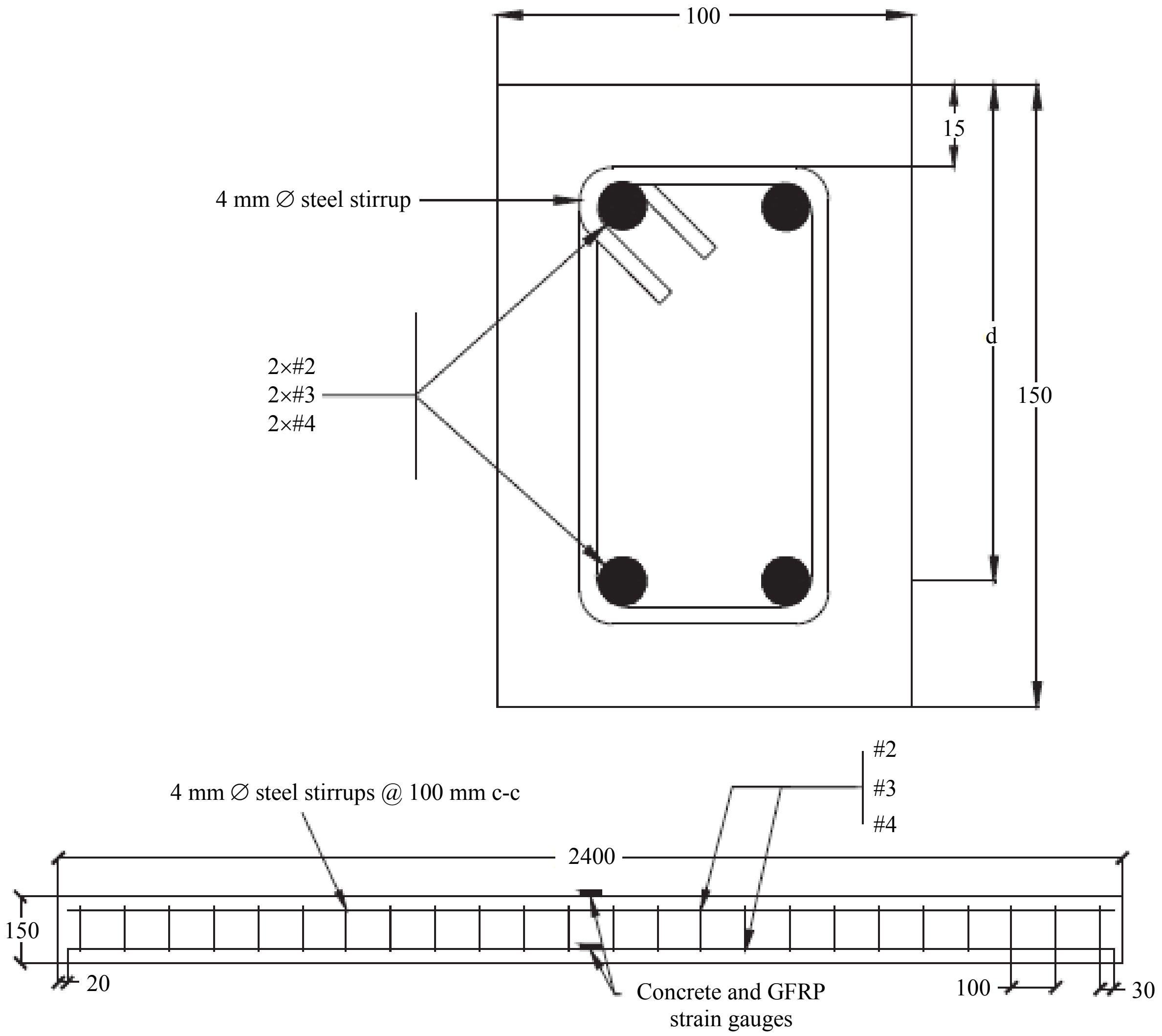

**Fig. 26:** Details of GFRP RC beams (Goldston *et al*., 2016)

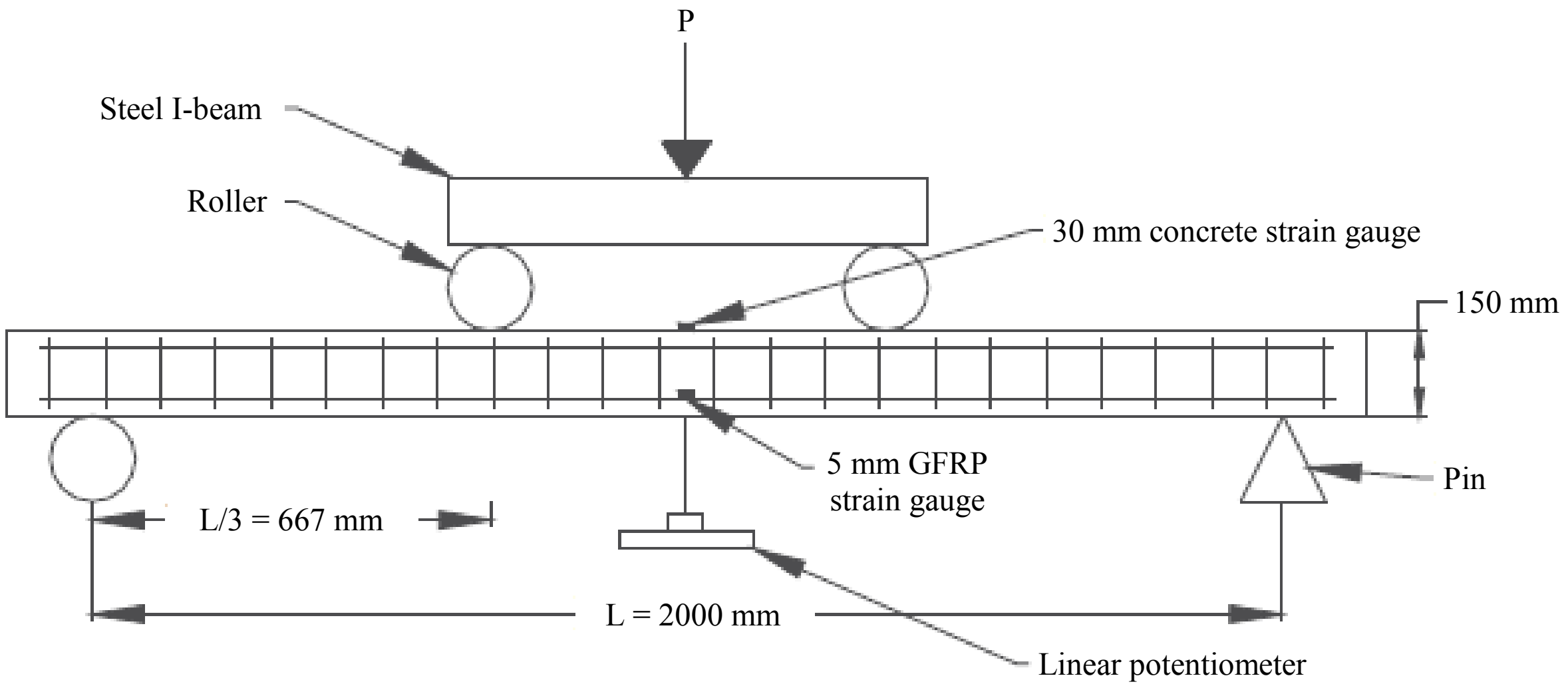

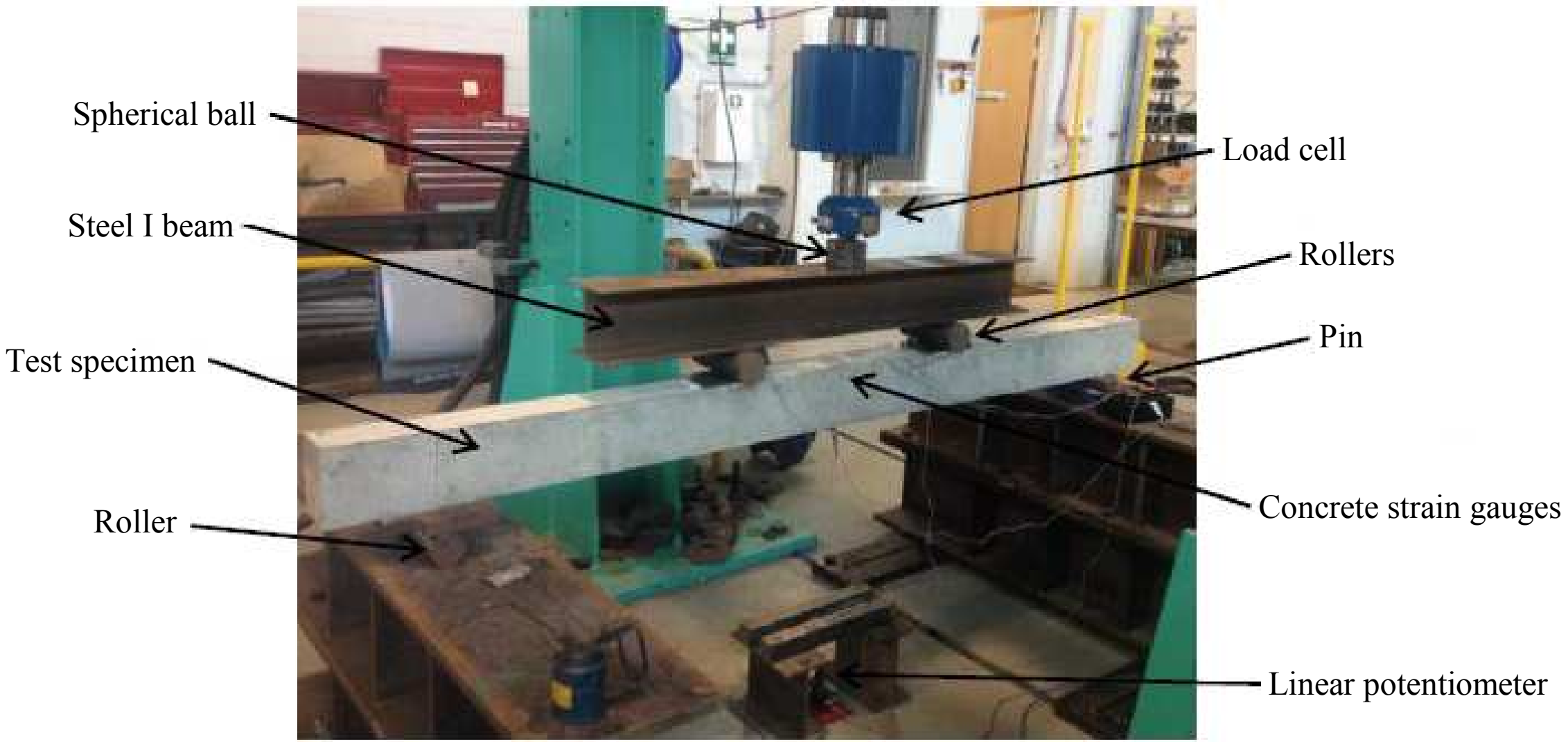

**Fig. 27:** Details of loading condition of RC beams (Goldston *et al*., 2016)

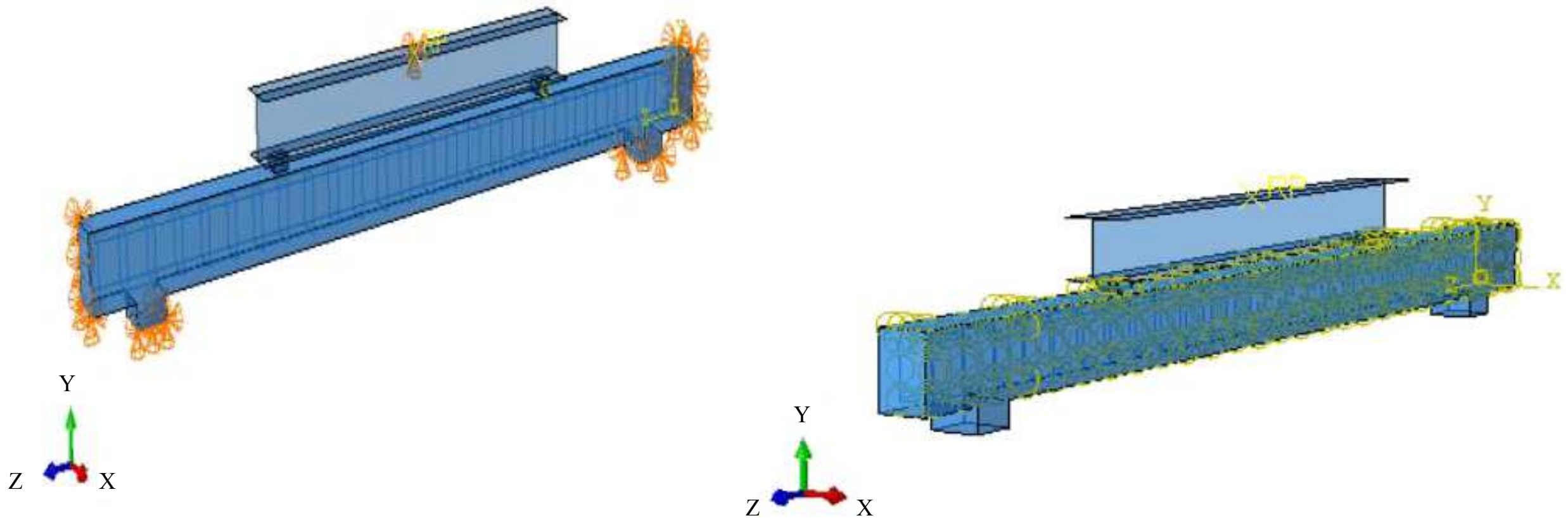

**Fig. 28:** Modeling of GFRP RC beam

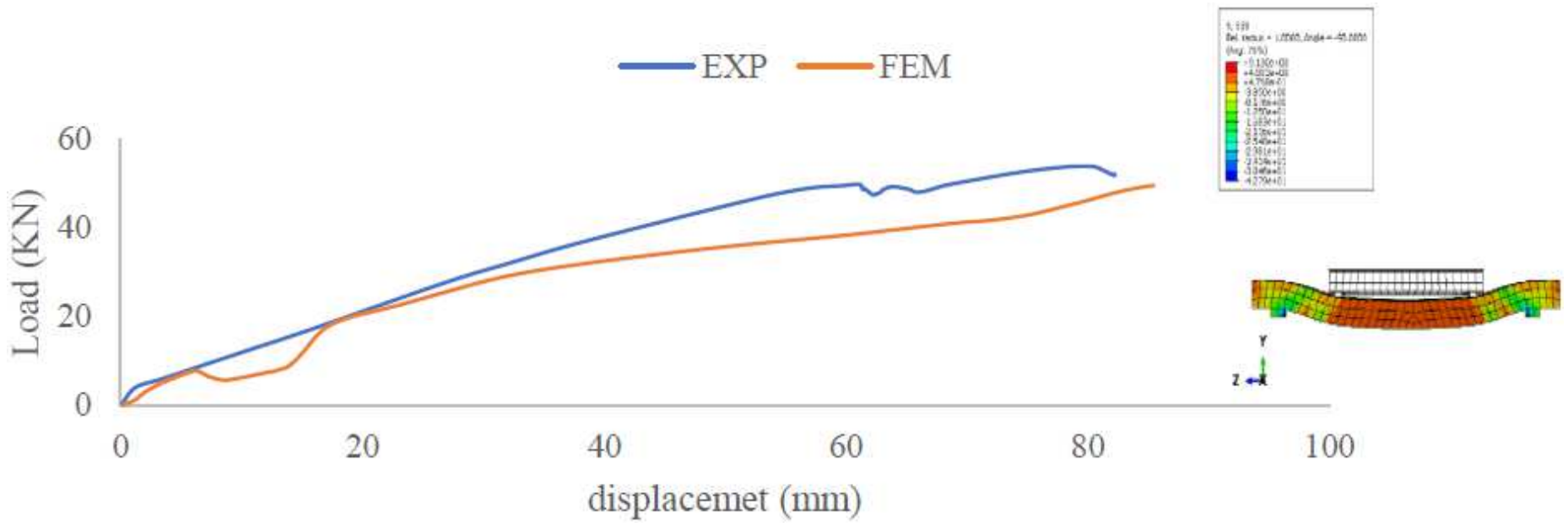

**Fig. 29:** Comparison between ABAQUS and Experimental results

The output of the finite element modeling versus experimental result is shown at Fig. 29. Considering the maximum base shear and displacement, the difference between the experimental and software's result is acceptable. The maximum displacement in ABAQUS is 85.43 millimeter representing only 3.8% difference from the experimental output which was 82.3 millimeter. Also, the analytical maximum shear base force was determined as 49.58 KN which is 7.8% lower than the experimental value of 53.78 KN. Figures 30 and 31 illustrate the evaluation of the load and displacement for a variety of reinforced concrete beams and reinforced composite rebar with impact loading at different drop height.

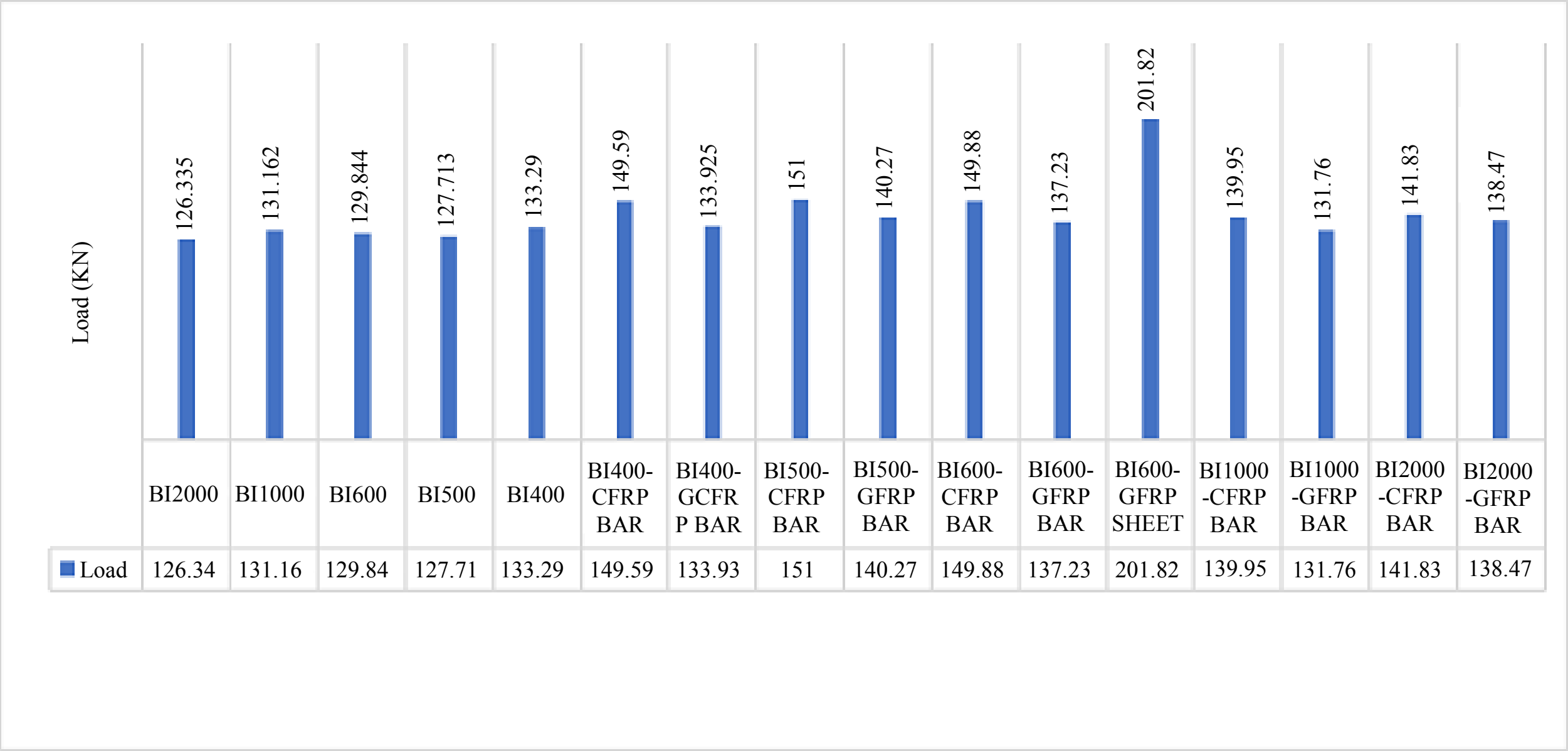

**Fig. 30:** Loads of BI specimens subjected to impact loadings at different heights

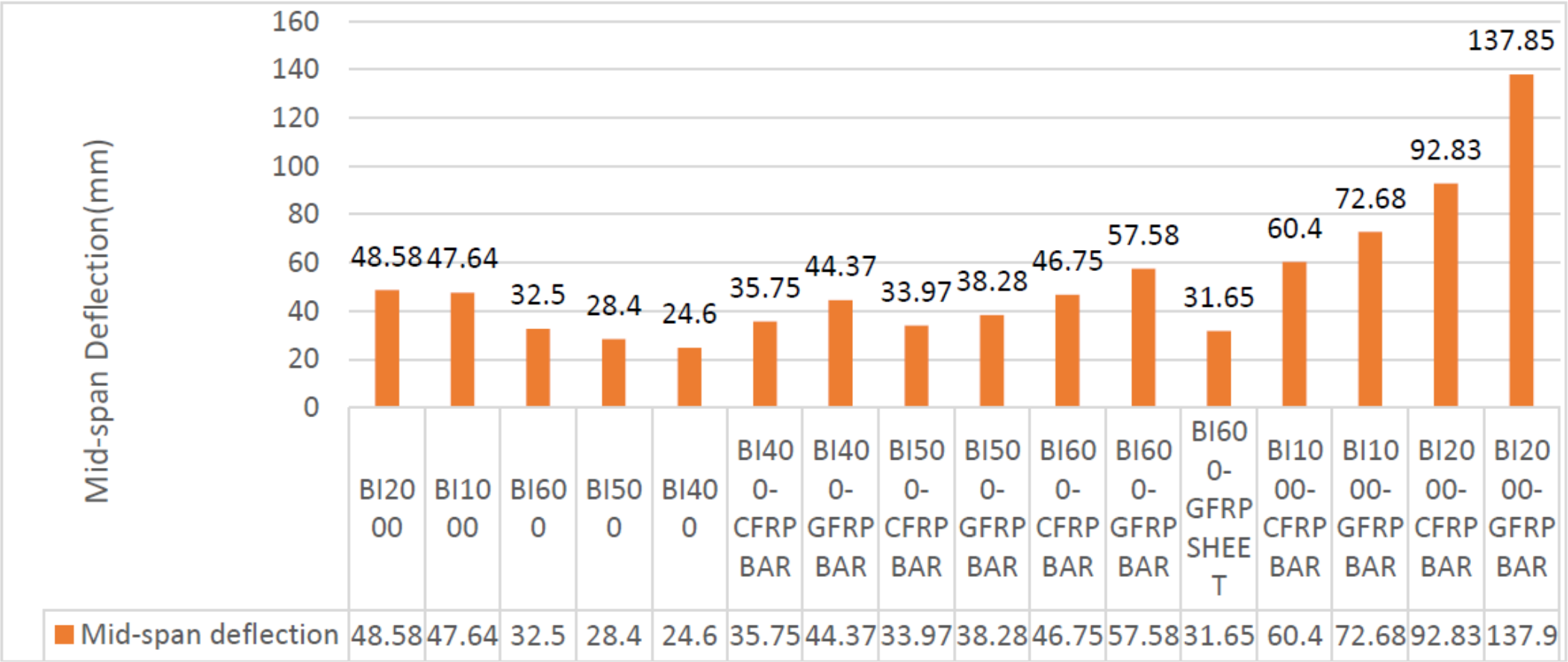

**Fig. 31:** Displacements of BI specimens in impact of varying heights

Investigating the loads in Fig. 30 and consider specimens BI of quasi-static load, specimen BI-400 illustrates the largest load capacity but the shortest throw height. Figure 31 shows the mid span deviation (displacement) at different throw heights. As shown, displacement increases with the height of the drop. Also, glass rebar increases the displacement while adding carbon rebar can increase the capacity. The highest increase in bearing related to the use of carbon rebar samples are BI500, the highest displacement (ductility) BI2000 reinforced with glass rebar.

Again, considering the load-displacement diagrams (deviation mid span beam) of Fig. 30 and 31 and comparing the unreinforced specimen under quasi-static load with the glass fibers reinforced one, one can see that the load capacity of sample BI600-GFRP is higher because of the external strengthening. The experimental results of the BS-GFRP beam strengthened by glass fiber show 29.3% increase in bearing capacity, while the analytical results show 30.03% increase. Also, BI-600-GFRP beam show an increase in bearing capacity of about 120.15% compare to the first sample. The corresponding analytical increase is 201.81%. A comparison between samples under quasi-static loads without and with GFRP and CFRP reinforcement show that the increase in base shear (bearing capacity) is 45.05% and the increase in displacement is 12.01% for CFRP sample. Also, GFRP sample leads to an increase in base shear amount of 39.22% and displacement of 28.96%. This indicates that using CFRP rebar in reinforced concrete beam under quasi-static load would increase bearing capacity and decrease displacement compare to GFRP rebar.

**Table 7:** Comparison between numerical modeling of reinforced and non-reinforced

| Difference displacement (%) | Difference base shear forces (%) | Specimen |
|---|---|---|
| 12.01 | 45.05 | BS-CFRP BAR |
| 28.96 | 39.22 | BS-GFRP BAR |
| 31.16 | 10.89 | BI400- CFRP BAR |
| 44.56 | 0.47 | BI400- GFRP BAR |
| 16.4 | 15.42 | BI500-CFRP BAR |
| 25.8 | 8.95 | BI500-GFRP BAR |
| 30.48 | 13.37 | BI600-CFRP BAR |
| 43.56 | 5.38 | BI600-GFRP BAR |
| 21.1 | 6.28 | BI1000-CFRP BAR |
| 34.44 | 0.45 | BI1000-GFRP BAR |
| 53.42 | 10.92 | BI2000-CFRP BAR |
| 64.75 | 8.75 | BI2000-GFRP BAR |

A comparison of samples under impact loading show that all samples reinforced with CFRP rebar have higher bearing capacity than that of GFRP rebar specimens, while the displacement in specimens containing glass rebar were far more than carbon. BI2000-CFRP Bar shows increase in shear base rate of 10.92% and BI 2000-GFRP Bar rate of 8.75%, as well as displacement 53.42 and 64.75% respectively. Summary of the above results are tabulated in Table 7.

## Conclusion

In this study, the finite element software, ABAQUS, was used to analytically investigate the behavior of concrete beams reinforced with carbon, glass, steel bars and GFRP sheets and subjected to different dynamic loading conditions (quasi-static, impact). Based on the analytical results and experimental verifications, the following conclusions can be drawn:

- Results of the finite element model using ABAQUS show good agreements with the experimental results
- In case of impact loadings, the load capacity of specimens reinforced with GFRP sheet were much higher than that of streel or CFRP and GFRP bars. On the other hand, the midpoint
- Deflection of beam for GFRP bar is higher than other beams
- By increasing the drop height of the hammer, the load capacity is decreased but midpoint deflection is increased. While CFRP bars improved the load capacity, GPRP bars improved ductility
- Concrete Beams reinforced with CFRP bars have higher quasi-static load capacity than that with GFRP bars

## Acknowledgment

The authors would like to thank their colleagues for the continuous support and contributions. We also would like to thank the anonymous reviewers very much whose useful comments and suggestions have helped strengthen the content and quality of this paper.

## Author's Contributions

**Sajjad Roudsari, Sayed Soleimani and Mona HaghighiFar:** Performed laboratory experiments, Numerical Analysis and conducted data analysis of the research. Also, participated in writing the manuscript.

**Sameer Hamoush and Taher Abu-Lebdeh:** Provided the research topic and guided the research development, experimental plan and data analysis. Also, participated in writing the manuscript.

## Ethics

This article is an original research paper. There are no ethical issues that may arise after the publication of this manuscript.